\documentclass[a4paper,fleqn]{scrartcl}
\usepackage[latin1]{inputenc}
\usepackage[T1]{fontenc}
\usepackage{lmodern}
\usepackage[pdftex]{graphicx}
\usepackage{dcolumn}
\usepackage{bm}
\usepackage{amsmath}
\usepackage{caption}[2008/07/24]
\newcommand{\bra}{\left\langle A0\right|}
\newcommand{\ket}{\left| A0\right\rangle}
\newcommand{\bran}{\left\langle A+2\right|}
\newcommand{\ketn}{\left| A+2\right\rangle}
\newcommand{\bram}{\left\langle A\right|}
\newcommand{\ketm}{\left| A\right\rangle}
\begin{document}

\title{Migdal and his theory in J\"ulich.}
\author{J. Speth, F. Grümmer and S. Krewald, \\
        Institut für Kernphysik, Forschungszentrum Jülich, \\
				52425 Jülich, Germany}
\maketitle
\emph{Dedicated to the memory of A.B. Migdal on the occasion of his $100^{th}$ anniversary.}
\begin{abstract}
We review the application of Migdal's \emph{Theory of Finite Fermi System}  
to the structure of deformed nuclei, approaches beyond the conventional linear response, and microscopic calculations of the \emph{Migdal-parameters.}
\end{abstract}
\section{Personal recollection by Josef Speth}
In 1975 I met Arkadi Benediktowitsch for the first time in Gerry Brown's institute in Stony Brook. He was interested in our numerical results of his theory of finite Fermi systems. During a visit in Dubna in 1977 he invited me to his apartment in Moscow for lunch and discussions with some of his collaborators. He visited  our institute in J\"ulich 1980 for a month, where he impressed not only physicists but also non-physicists at various parties in J\"ulich. The photograph shows him as a \emph{magician} who makes some conjuring trick for my children.

\begin{figure}[htbp]
\begin{center}
\includegraphics[width=6cm]{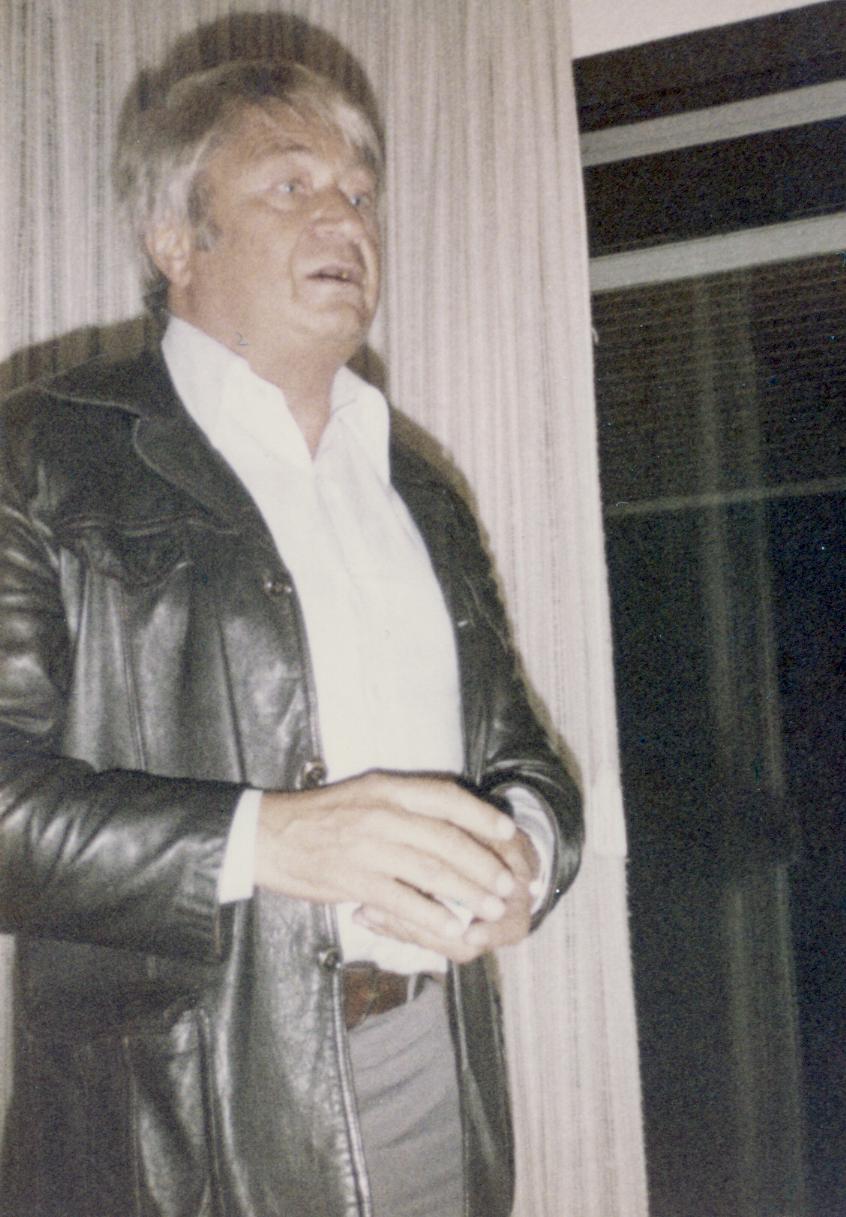}
\end{center}
\caption{\label{fig:Migdal}
A.B. Migdal as \emph{magician} in Speth's house (1980).}
\end{figure}

\section{Introduction}
There exist several reviews on Migdal's \emph{Theory Finite Fermi System} \cite{Migdal67} (TFFS) where the application of the theory to various aspects of atomic nuclei has been presented. The original approach is restricted to the solution of the linear response function, which is connected with the 4-point many-body Green function (GF). In the early review by Speth et al. \cite {Speth77} numerical results for nuclei in the lead region have been summarized. They also reported on an extension to the 6-point GF, that had been previously developed in Ref. \cite{Speth70,Ring74}. Khodel and Saperstein investigated in their publications, which have been summarized in Ref. \cite{KS82}, a self-consistent version of the \emph{TFFS} and presented results for several closed shell nuclei. Dro\.zd\.z et al. \cite{Stan90} developed an important extension of the \emph{TFFS} which allows to describe giant resonances in a quantitative way. The authors include  1p1h and 2p2h configurations in a consistent manner and demonstrate the power of their method for electric, magnetic and charge exchange resonances. In a more recent work by Kamerdzhiev et al. \cite{rev04} another extension of the TFFS was reviewed. Here  $1p1h\otimes$phonon configurations were included in the conventional linear response equation, their effect on collective electric resonances discussed and results for closed shell nuclei presented. Finally Tselyaev \cite{Tselyaev89,Tselyaev07} presented a model where he included not only two phonon states but also pairing correlations in a consistent manner, which allows applications to open shell nuclei. With the exception of Ref. \cite{Migdal67,Speth77,Stan90} mostly electric properties of spherical nuclei have been discussed. Recent applications to neutron rich nuclei are reviewed in Ref. 
\cite{GS06,Krewald09,TsSp07,Ring04,LiTselyaev07,Ring01,Lenske02,Ring02}.

In the present review we demonstrate that the \emph{TFFS} has been successfully applied to strongly deformed nuclei where electric as well as magnetic properties in the rare earth and actinide region have been calculated. It is very important to mention that the derivation of the basic formulas within the many-body GF formalism is very general and the corresponding equations are not restricted to the \emph{TFFS}. We will see in the next section that for the numerical application one needs single particle energies, single particle wave functions and an effective (ph)-interaction. If pairing is included one also needs an effective particle-particle (pp) interaction. In the \emph{TFFS} the single particle quantities are deduced from phenomenological single-particle models and the residual interactions are parametrized. A comparison between the phenomenologically determined parameters and calculated ones from first principles can be found in Ref. \cite {Krew88}. In the self consistent approaches the single particle data are obtained from a mean field calculated with an effective Hamiltonian, Lagrangian or energy functional. The residual ph-interaction is given in such approaches as second derivative of the energy. In this review we want to show that the \emph{TFFS} is not only able to reproduce known data but that most of the predictions have been experimentally verified. The review may also give a guide line for those scientists which start to investigate deformed nuclei, though in different models and theories, respectively.
  
  In section 2 we define the GF and the corresponding equations of motion. These quantities are derived with the functional method developed by Kadanoff and Baym \cite{Baym62} and its generalization to systems with pairing proposed by Brenig and Wagner \cite{Wagner63}. In the derivation of the basic formulas we do not give the details but we outline the basic idea and refer to other reviews or the original publications for details. In this section we also define the second-order response function which has been successfully used to calculate isomer shifts of rotational states in the rare earth region \cite{Meyer73}. In section 3 we present results for deformed nuclei. This involves electric as well as magnetic low-lying and high-lying resonances. The microscopic results are compared with the well known phenomenological models. We discuss in sections 4 an extension of Migdal's linear response theory, the second order response. This formalism was application to a very specific nuclear structure effect, namely the isomer shifts of rotational states in deformed nuclei. These extremely small effects were measured with the help of the M\"ossbauer effect. In section 4 we review magnetic states in deformed nuclei. In section 5 we discuss microscopic calculations of the Migdal parameters from first principles and in section 6 a short summery is given. We refer mostly to the work done in Munich and J\"ulich, with additional references to the work by Urin \cite{Urin1,Urin2}, who investigated collective vibrations in deformed nuclei within the \emph{TFFS} in a different way and the application of the \emph{TFFS} to the Inglis cranking model by Birbrair \cite{Bir73}.

\section{Method}

\subsection{Many Body Green Functions}

The GF from which we start are defined as the ground-state expectation value of a N-particle system of the time order product of pairs of quasi-particle creation and annihilation operators. In the present context, these Green functions are functions of an external field $q(1,2)$, which is a formal device to derive higher GF and their equations of motion \cite{Baym62}. After one has performed the derivations one can put the source field equal to zero and arrives at the conventional definition. In a single particle basis the one-particle GF is given as:
 
\begin{equation}\label{eq:1}
g_q{(1,2)} = \frac{i}{\bra{T{\left\{U\right\}}\ket}}
\bra{T{\left\{U{a_{\nu_1}{(t_1)}{a_{\nu_2}^+{(t_2)}}}\right\}}}\ket
\end{equation}
with 
\begin{equation}\label{eq:2}
U=e^{i\int{d5d6\;q(5,6){a_{\nu_5}^+{(t_5)}{a_{\nu_6}{(t_6)}}}}}
\end{equation}
and the short hand notation:
\begin{equation}\label{eq:3}
\sum_{\nu_1}\int{dt_1} = \int{d1}.
\end{equation}
The two-particle and three-particle GF have correspondingly the form (with q(1,2)=0):
\begin{eqnarray}\label{eq:4}
g(13,24) &= (i^2)\bra T a_{\nu_1}{(t_1)}a_{\nu_3}{(t_3)}a_{\nu_2}^+{(t_2)}a_{\nu_4}^+{(t_4)}\ket \\
g(135,246) &= (i^3)\bra{T{ a_{\nu_1}{(t_1)}a_{\nu_3}{(t_3)}a_{\nu_5}{(t_5)}a_{\nu_2}^+{(t_2)}a_{\nu_4}^+{(t_4)}a_{\nu_6}^+{(t_6)}}}\ket .
\end{eqnarray}
The equation of motion for the one-particle GF is the Dyson equation which has the form:

\begin{equation}\label{eq:5}
\frac{i}{2}\int{d3\left\{S(1,3)\;+\;q(1,3)\;+\;\Sigma_q{(1,3)}\right\}g_q{(3,2)}}=\delta (1,2).
\end{equation}
with the abbreviations:
\begin{eqnarray}\label{eq:6}
\delta_{\nu_1\nu_2}\delta(t_1-t_2)\left\{i\frac{\delta}{\delta{t_1}}-\epsilon_{\nu_1}^0\right\} &= S(1,2).
\end{eqnarray}
The quantity $\Sigma$ is the \emph{self-energy or mass operator}, an effective one-body potential, which in principle is given by the bare interaction of the corresponding Hamiltonian and the two-particle GF \cite{Speth77}. We define the linear response function $L$ by the functional derivative due to the field $q(1,2)$ of the one-particle GF:
\begin{equation}\label{eq:7}
\frac{\delta g_q{(1,2)}}{\delta q(4,3)}\;=\;L(13,24)\;=\;g(13,24)\;-\;g(1,2)g(3,4). 
\end{equation}
The functional derivative of the Dyson equation gives an integral equation for the response function
\begin{eqnarray}\label{eq:8}
L(13,24)= -g(1,4)g(3,2) -i\int{d5d6d7d8\;g(1,5)K(57,68)L(83,74)g(6,2)},
\end{eqnarray}
where we introduced the effective two-body interaction $K$ via
\begin{equation}\label{eq:8a}
\frac{\delta \Sigma (1,2)}{\delta g(3,4)}=iK(13,24).
\end{equation}
The change of the one-particle GF $\delta g_q$ due to an external field $\delta q$ is given in linear response :
\begin{eqnarray}\label{eq:8b}
\delta g_q(1,2)= \int{ d3d4 L(13,24)\delta q(4,3)}.
\end{eqnarray}
Analogously to the linear response function one defines the second order response function as:
\begin{eqnarray}\label{eq:9}
\frac{\delta^2 g_q{(1,2)}}{\delta q(4,3)\delta q(6,5) }\;=\;L(135,246)  \;=\; g(135,246) +2g(1,2)g(3,4)g(5,6) \\ \nonumber
-g(13,24)g(5,6) -g(15,26)g(3,4)-g(35,46)g(1,3).
\end{eqnarray}
As we will see in the following Eq.(\ref{eq:8}) is the central equation in the conventional \emph{TFFS}. The one-particle GF and the two-particle GF are defined self-consistently by a system of non-linear equations. This is, however, of little use for practical applications. In order to arrive at solvable equations, one applies Landau`s quasi particle concept and his renormalization procedure, which he developed for his \emph{Theory of Fermi Liquid} \cite{Landau}. Migdal introduced this concept into the nuclear many-body problem. Here, the quasi particles are the single-particle states. Following Landau, one splits the (Fourier transformed) one-particle GF into a pole part and a remainder.  Written in the configuration space of the single-particle wave functions ${\varphi_\nu}$ the equation has the form: 
 
\begin{eqnarray}\label{eq:10}
g_{\nu_1 \nu_2}(\epsilon) = z_{\nu_1}\frac{\delta_{\nu_1 \nu_2}}{ \epsilon_{\nu_1} - \epsilon + i\eta \; \textrm{sign}(\epsilon_{\nu_1} -\mu)}
                      + \;g^{r}_{\nu_1 \nu_2}(\epsilon).\;\;\;\;\;\;\;
\end{eqnarray}
The $\epsilon_{\nu}$ are the single-particle energies, $z_{\nu}$ the single particle strength and  $\mu$ is the Fermi energy. 
 The main goal is to obtain an equation for the response function that can be solve in practice. With the ansatz in Eq. (\ref{eq:10})
one writes the product of two GF as a singular part $S$ and a rest $B$:
 \begin{eqnarray}\label{eq:11}
 g(\epsilon,\Omega)g(\epsilon)\;=\;S(\epsilon,\Omega)\;+\;B(\epsilon,\Omega).
 \end{eqnarray}
 The singular part has the form:
  \begin{eqnarray}\label{eq:12}
  A_{\nu_1\nu_2,\nu_3\nu_4}(\epsilon,\Omega)\;= \;
  2\pi\imath z_{\nu_1}z_{\nu_2}\delta_{\nu_1\nu_3}\delta_{\nu_2\nu_4}\frac{n_{\nu_1}-n_{\nu_2}}                   
{\epsilon_{\nu_1}-\epsilon_{\nu_2}-\Omega}
\delta(\Omega-
\frac{\epsilon_{\nu_1}+\epsilon_{\nu_2}}{2})                                 
\end{eqnarray}
here the $n_{\nu}$ are the occupation numbers for quasi particles: $1$ and $0$ for particles and holes, respectively.

\subsection{Linear response}

From the linear response equation (\ref{eq:8},\ref{eq:8b}) on can derive an equation for the change of the density $\delta \rho$ due to an external field $\delta q$:
\begin{eqnarray}\label{eq:13a}
\left( \epsilon_{\nu_1}-\epsilon_{\nu_2}- \Omega\right)\delta \rho_{\nu_1 \nu_2} \; = \;
\left(n_{\nu_1}-n_{\nu_2}\right) (\delta {\widetilde{q}}_{\nu_1 \nu_2}(\Omega)+\sum_{\nu_3 \nu_4}F^{ph}_{\nu_1 \nu_4 \nu_2 \nu_3} \;\delta \rho_{\nu_3 \nu_4}).
\end{eqnarray}
Here $F^{ph}$ and $\delta \tilde {q}$ are the renormalized \emph{ph}-interaction and renormalized external field, respectively, which includes
the $z_\nu$ factors as well as the \emph{regular} part $B$ of Eq.(\ref{eq:11}).
The homogeneous part of Eq.(\ref{eq:13a}) is formally identical with the conventional \emph{Random Phase Approximation} (RPA)
\begin{eqnarray}\label{eq:13}
\left( \epsilon_{\nu_1}-\epsilon_{\nu_2}- \Omega\right)\chi^{m}_{\nu_1 \nu_2} = \;
\left(n_{\nu_1}-n_{\nu_2}\right)\sum_{\nu_3 \nu_4}F^{ph}_{\nu_1 \nu_4 \nu_2 \nu_3} \;\chi^{m}_{\nu_3 \nu_4}.
\end{eqnarray}
From Eq.(\ref{eq:13}) one calculates excitation energy $\Omega_m$  and the corresponding (renormalized) ph-transition amplitude $\chi^{m}_{\nu_1 \nu_2}$ from the ground-state to the excited state $\left|Am\right\rangle$ of an even-even nucleus with mass number A. From the latter ones follow the expectation values of one-body operators:
\begin{eqnarray}\label{eq:14}
\left\langle Am \right|Q\left|A0\right\rangle = \sum_{\nu_1 \nu_2} Q^{eff}_{\nu_1 \nu_2}\chi^{m}_{\nu_2 \nu_1}.
\end{eqnarray}
The renormalized single-particle operators $Q^{eff}$ also include  the $z_\nu$ factors and the \emph{regular} part $B$. In the case of electric multipole operators $Q^{eff}$ correspond to the bare operators due to charge conservation, whereas the magnetic operators are parametrized, with universal parameters \cite{Migdal67,Speth77}. 
The response function includes also an equation for moments and transitions in the neighboring odd-mass nuclei \cite{Migdal67,Speth77},
\begin{eqnarray}\label{eq:15}
\left\langle A\pm 1,\alpha \right|Q\left|A\pm 1,\beta\right\rangle 
\;-\; \delta_{\alpha,\beta}\left\langle A0 
\right|Q\left|A0\right\rangle  \; = \; 
\tau_{\alpha\beta}(\epsilon_{\alpha\beta},Q),
\end{eqnarray}
the corresponding equation for the vertex operator \\ $\tau_{\alpha\beta}(\epsilon_{\alpha\beta},Q)$ has the form:
\begin{eqnarray}\label{eq:16}
\tau_{\nu_1\nu_2}(\epsilon_{\alpha\beta},Q)\;=\;Q^{eff}_{\alpha\beta}\delta_{\nu_1\alpha}\delta_{\nu_2\beta}\;
+\;\sum_{\nu_3 \nu_4}F^{ph}_{\nu_1 \nu_4 \nu_2 \nu_3}\frac{n_{\nu_3}-n_{\nu_4}}{\epsilon_{\nu_3}-\epsilon_{\nu_4}-\epsilon_{\alpha\beta}}
\tau_{\nu_3\nu_4}(\epsilon_{\alpha\beta},Q).
\end{eqnarray}

Here $\epsilon_{\alpha\beta}$ is the energy difference between the two states $\alpha$ and $\beta$. From Eq.(\ref{eq:15}) we see that in the case of moments one can only calculate the difference between the even and odd mass nuclei. This, however, allows a very precise calculation of the differences of charge distributions (isotope shifts).

\subsection{Many Body Green functions including pairing correlations}

With the exception of closed shell nuclei, pairing correlations play an important role in nuclei. Therefore one needs an extension of the previously discussed \emph{Theory of Finite Fermi systems} which includes pairing correlations.
Such an extension has first been presented by Larkin and Migdal \cite{Larkin63}. Here we give a more general derivation \cite{Birbrair,Sergey69,Speth69} which is based on generalized GFs \cite{Gorkov}. We introduce a one-particle GF matrix of the form:

\begin{align}\label{eq:17}
G_{\kappa \lambda}^{kl}\;=\;
&\left(
\begin{array}{ll}
G_{\kappa \lambda}^{11}\;\;G_{\kappa \lambda}^{12}  \\ \\
G_{\kappa \lambda}^{21} \;\;G_{\kappa \lambda}^{22} 
\end{array}
\right) \; = \;
 \frac{i}{\bram{T{\left\{U\right\}}\ketm}} \; \times \\
&\left(
\begin{array}{ll}
\bram{T{\left\{U{a_{\kappa}a_{\lambda}^+}\right\}}}\ketm & \;\;-\bram{T{\left\{U{a_{\kappa}{a_{\overline{\lambda}}}}\right\}}}\ketn\ \\ 
\bran{T{\left\{U{a_{\overline{\kappa}}^+{a_{\lambda}^+}}\right\}}}\ketm & \;\;-\bran{T{\left\{U{a_{\overline{\kappa}}^+{a_{\overline{\lambda}}}}\right\}}}\ketn \;	
\end{array}
\right) \nonumber
\end{align}

The bar on the indices denotes the time-reversed states. The two-particle GFs and the response functions are defined in the same way as before and are obtained as functional derivatives. The generalized Dyson equation possesses also four components which includes four mass operators. By functional derivation of the Dyson equation one obtains the integral equations for the four response functions. The ansatz for the (diagonal) pole parts of the four  one-particle GFs can be written in a compact form \cite{Meyer73}:

\begin{eqnarray}\label{eq:19}
G^{(0)}_{\lambda}(\omega)\;=\;-\left(\frac{{L_{\lambda}}}{\omega+E_{\lambda}-i \eta} +\frac{T_{\lambda}}{\omega-E_{\lambda}+i\eta}\right)
\end{eqnarray}

\begin{equation}\label{eq:20}
L_{\lambda}\;=\;
\left(
\begin{array}{ll}
v^{2}_{\lambda}  \;\; -u_{\lambda}v_{\lambda} \\ 
u_{\lambda}v_{\lambda }\;\; -u^{2}_{\lambda} 
\end{array}
\right) 
  \; ; \;
T_{\lambda}\;=\;
\left(
\begin{array}{ll}
\;\;u^{2}_{\lambda}  \;\;\;\;\; u_{\lambda}v_{\lambda}\\ 
-u_{\lambda}v_{\lambda} \;\; -v^{2}_{\lambda} 
\end{array}
\right) 
\end{equation}
with the BCS quantities:
\begin{equation}\label{eq:22}
v^{2}_\lambda =\frac{1}{2}\left(1-\frac{\epsilon_{\lambda}-\mu}{E_{\lambda}}\right)\;\;\;\;u^{2}_{\lambda}=\frac{1}{2}\left(1+\frac{\epsilon_{\lambda}-\mu}{E_{\lambda}}\right)
\;\;\;\;
E^{2}_{\lambda}=\left(\epsilon_{\lambda}-\mu\right)^2+\Delta^2_{\lambda}.
\end{equation}
The gap $\Delta$ is given by the usual gap equation:
\begin{equation}\label{eq:24}
\Delta_{\lambda}=-\sum_{\kappa} F^{pp}_{\overline{{\lambda}}{\lambda},\overline{\kappa}{\kappa}}\frac{\Delta_\kappa}{2E_\kappa}
\end{equation}
here $F^{pp}$ is the renormalized particle-particle (pp) interaction, which also enters in the equation for the response function.

\subsection{Quasi-particle RPA}

The four coupled equations for the response functions can be reduced to two coupled equations. With the ansatz for the pole part of the one-particle GFs given in Eq.(\ref{eq:19}) plus a \emph{regular} part one performs an analog renormalization procedure as described in the previous section. The final equations have the form of the well known quasi-particle RPA (QRPA) equations which allow to calculate e.g. collective excitations in super fluid Fermi systems of two quasi-particle type. These equations have been previously derived in different ways by Bogoliubov \cite{Bogolyubov59}, Baranger \cite{Baranger60} and Belyaev \cite{Belyaev}. Birbrair \cite{Birbrair} and Kamerdzhiev \cite{Sergey69} also used the GF formalism making explicit the difference between pp- and ph- interaction. Here we give the compact form of the equations derived by Baranger \cite{Baranger60}: 

\begin{eqnarray}\label{eq:25}
\left(E_\lambda + E_\kappa\right)Z^+_{\lambda \kappa}+\sum_{\nu \mu}\left(\eta^{+}_{\lambda \kappa}F^{ph+}_{\lambda
 \mu, \kappa \nu} \eta^{+}_{\mu \nu} \right. 
\left. +\xi^{+}_{\lambda \kappa}F^{pp+}_{\lambda \mu, \kappa \nu}\xi^+_{\mu \nu}\right)Z^-_{\mu \nu}
=\Omega Z^-_{\lambda \kappa} 
\end{eqnarray}

\begin{eqnarray}\label{eq:26}
\left(E_\lambda + E_\kappa\right)Z^-_{\lambda \kappa}+\sum_{\nu \mu}\left(\eta^{-}_{\lambda \kappa}F^{ph-}_{\lambda \mu, \kappa \nu} 
\eta^{-}_{\mu \nu} \right. 
\left.+\xi^{-}_{\lambda \kappa}F^{pp+}_{\lambda \mu, \kappa \nu}\xi^-_{\mu \nu}\right)Z^+_{\mu \nu}=  
\Omega Z^-_{\lambda \kappa} 
\end{eqnarray}

with the normalization condition:
\begin{equation}\label{eq:27}
2\sum_{\lambda \kappa}Z^+_{\lambda \kappa}\left(Z^-_{\lambda \kappa}\right)^*=1.
\end{equation}

The various quantities in Eqs.(\ref{eq:25}) and (\ref{eq:26}) are defined as follows:

\begin{equation}\label{28}
 \xi^{\pm}_{\lambda \kappa}=u_\lambda u_\kappa\mp v_\lambda v_\kappa;  \;\;\;  \;\;  \eta^{\pm}_{\lambda \kappa}=u_\lambda v_\kappa\mp v_\lambda u_\kappa ,
\end{equation}
 
and 
 
\begin{equation}\label{29}
 F^{ph \pm}_{\lambda\mu, \kappa \nu} =\frac{1}{2}\left(F^{ph}_{\lambda\mu, \kappa \nu} \pm F^{ph}_{\lambda  \overline{\nu}, \kappa \overline{\mu}}\right).
\end{equation} 

From the solutions of Eqs. (\ref{eq:25}) and (\ref{eq:26}) we obtain the excitation energies $\Omega $ and the  amplitudes $Z^\pm$ which are connected with the transition probabilities $B(EL)$ in the following way.
\begin{eqnarray}\label{eq:30}
B\left(EL\right)=e^2\left(2-\delta_{K0}\right)\left|\sum_{\lambda \kappa}\left(r^2Y_{LK}\right)_{\lambda \kappa}\chi^+_{\lambda \kappa}\right|^2.
\end{eqnarray}
with
\begin{equation}\label{31}
\chi^{\pm}_{\lambda \kappa}=\eta^{\pm}_{\lambda \kappa}Z^{\pm}_{\lambda \kappa};\;\;\;\;\chi^{\pm}_{\lambda \kappa} = \frac{1}{2}\left( \widetilde{\chi}^{0m}_{\lambda \kappa}\pm \widetilde{\chi}^{0m}_{\overline{\lambda}\overline{\kappa}}\right),
\end{equation}
where $\widetilde{\chi}^{0m}$ is the renormalized version of the following (unrenormalized) matrix element of an A-particle system:
\begin{equation}\label{32}
\chi^{0m}_{\lambda \kappa}=\left\langle A0 \right|a^+_\lambda a_\kappa \left|Am\right\rangle
\end{equation}
Eq. (\ref{eq:30}) is the analog to Eq. (\ref{eq:14}) for super fluid systems. Like in the non super fluid case, the electric multipole operators can be replaced by the bare operators, because of charge conservation. In the magnetic case one has to use renormalized operators. The analog to Eq. (\ref{eq:16}) (moments and transitions in odd mass nuclei) is given in Ref. \cite{MS72}. This equation has been successfully applied to the calculation of isomer shift \cite{Speth69} and isotope shifts \cite{MS72} in deformed odd mass nuclei.

\subsection{Second order response theory}

Within the linear response theory one calculates the change of the expectation value of a single particle operator in an external field. Due to the linear relation, the single particle operators and the external field have to have the same operator structure. 
One of the especially nice applications of the following extended version of the \emph{TFFS} is the calculation of the change of the nuclear charge radii in rotational states. These tiny effects were measured with help of the M\"ossbauer effects. Here one calculates within the cranking model the change of the scalar operators $ \delta r{_p}^2 $ due to the Coriolis perturbation ${\delta \bf{q}}=-\Omega_c\ \bf{J_x}$ which is a pseudo vector operator. Here $\Omega_c$ is the so called cranking parameter. 

The change of the density in linear response $\rho^{(1)}$ due to the Coriolis perturbation has the form \cite{Meyer73}:

\begin{eqnarray}\label{eq:32}
\rho^{(1)}_{\lambda,\kappa}=\frac{(\eta_{\lambda,\kappa}^{(-)})^2}{E_{\lambda,\kappa}}\left[(J_x)_{\lambda,\kappa}-\sum_{\mu,\nu}F^{ph}_{\lambda\mu, \kappa \nu}\rho^{(1)}_{\nu,\mu}\right]-
\frac{\eta_{\lambda,\kappa}^{(-)}\xi_{\lambda,\kappa}^{(-)}}{E_{\lambda,\kappa}}\sum_{\mu,\nu}F^{pp}_{\overline{\lambda} \kappa, \overline{\mu} \nu}\frac{\xi^{-}_{\mu \nu}}{\eta^-_{\mu \nu}}\rho^{(1)}_{\mu \nu}.
\end{eqnarray}
It is obvious that the change of the radius is zero in linear response. This equation has been previously derived by Migdal \cite{Migdal59} and Birbrair \cite{Bir68} and has been used to calculate moments of inertia and gyromagnetic ratios \cite{Meyer72}. The equation for the change of the density in second order $\rho^{(2)}$ has the same structure as Eq. (\ref{eq:32}). All quantities that enter in the equation are know from the linear response theory with the exception of an effective three particle interaction which has been neglected in all applications. 
\begin{eqnarray}\label{eq:33}
\rho^{(2)}_{\lambda,\kappa}= \widetilde{\rho}^{(2)}_{\lambda,\kappa}[\textrm{inh}]-\frac{(\eta_{\lambda,\kappa}^{(+)})^2}{E_{\lambda,\kappa}}\sum_{\mu,\nu}\widetilde{F}^{ph}_{\lambda\mu, \kappa \nu}\rho^{(2)}_{\nu,\mu}-
\frac{\eta_{\lambda,\kappa}^{(+)}\xi_{\lambda,\kappa}^{(+)}}{E_{\lambda,\kappa}}\sum_{\mu,\nu}\widetilde{F}^{pp}_{\overline{\lambda} \kappa, \overline{\mu} \nu}\frac{\xi^{+}_{\mu \nu}}{\eta^+_{\mu \nu}}\rho^{(2)}_{\mu \nu}.
\end{eqnarray}

The inhomogeneous term depends in a complicated way quadratically on the Coriolis perturbation. The very lengthy formula for $\widetilde{\rho}^{(2)}[\textrm{inh}]$  and the modified $\widetilde{F}^{ph}$ and $\widetilde{F}^{pp}$ are given in Ref. \cite{Meyer73}. 

\section{Application to deformed nuclei}

In order to solve the basic equations one needs as input single particle-wave functions, single-particle energies and the ph-and pp-interaction. Migdal has designed his theory in close connection to Landau's Fermi liquid theory. Therefore one takes as far as possible the input data from experiment or from models which reproduce the needed experimental data as good as possible. 
  If the RPA equation is derived within the many-body Green function formalism \cite{Speth77} one obtains an explicit form for the ph-interaction $F^{ph}$ that depends on the single-particle model and the size of the configuration space. In addition it is nonlocal and energy-dependent. In practice the ph-interaction is not calculated from that formula but it is parametrized. Following Landau's procedure $F^{ph}$ is transformed into momentum space and one considers the interaction on the Fermi surface of nuclear matter. Here one can replace the energies by the Fermi energy and the magnitude of the momenta by the Fermi momentum. In this approximation $F^{ph}$ depends only on the angle between the ph-momenta $\bf{P}$ and $\bf{P'}$ before and after the collision; (we suppress the spin and isospin dependence)

\begin{eqnarray}\label{eq:3a}
F^{ph}\left( \frac{ \mathbf{P}\cdot \mathbf{P'}} {P^{2}_F} \right) = \sum_{l=0}^{\infty} F_l P_l\left( \frac{ \mathbf{P}\cdot \mathbf{P'}} {P^{2}_F} \right).
\end{eqnarray}
Here $P_l(x)$ is the Legendre polynomial of order $l$ and the constants $F_l$ are the famous Landau-Migdal parameters. One introduces dimensionless parameters by defining:
\begin{eqnarray}\label{eq:4a}
F_l=C_0f_l.
\end{eqnarray}
Here $C_0 $ is the inverse density of states at the Fermi surface.
One choses density dependent parameters which correct for the finite size of the nuclei. 
For deformed nuclei an axially deformed Fermi distribution was used.

\begin{eqnarray}\label{eq:5a}
F^{ph}(1,2)=C_{0}\delta (\mathbf{r_1}-\mathbf{r_2})\cdot 
\left[f_0(\rho) + f^{\prime}_0(\rho)\mathbf{\tau_1} \cdot\mathbf{\tau_2} +g_0(\rho)\mathbf{\sigma_1} \cdot\mathbf{\sigma_2} +g^{\prime}_0(\rho) \mathbf{\sigma_1}\cdot\mathbf{\sigma_2}\mathbf{\tau_1} \cdot\mathbf{\tau_2}\right] 
\end{eqnarray}

The pp-interaction is treated in the same way. In all applications so far, the interaction is restricted to the scalar-isoscalar component, with parameters which are in some cases density dependent.

 The single-particle wave functions are taken from a single-particle model and the single-particle energies are taken as far as possible from experiment. The deformed rare earth nuclei and the actinides are well described within the unified model \cite{BM} that provides a good working single-particle model. Here the authors obtained the single particle wave functions from a deformed Woods-Saxon potential \cite{Vogeler}. The most critical question has been the choice of the single particle energies. In the case of the isomer shifts, which was the first application of the \emph{TFFS} to deformed nuclei, the level scheme by Ref. \cite{Gustaf} has been used with some corrections due to new experimental informations. For the solution of the QRPA equation which had been done some years later the theoretical level schemes have been thoroughly readjusted to new                       experimental data. \cite{Zawischa78}.
\section{Results}

\subsection{Classical quadrupole shape oscillations}

\begin{figure}[htbp]
\begin{center}
\includegraphics[width=12cm]{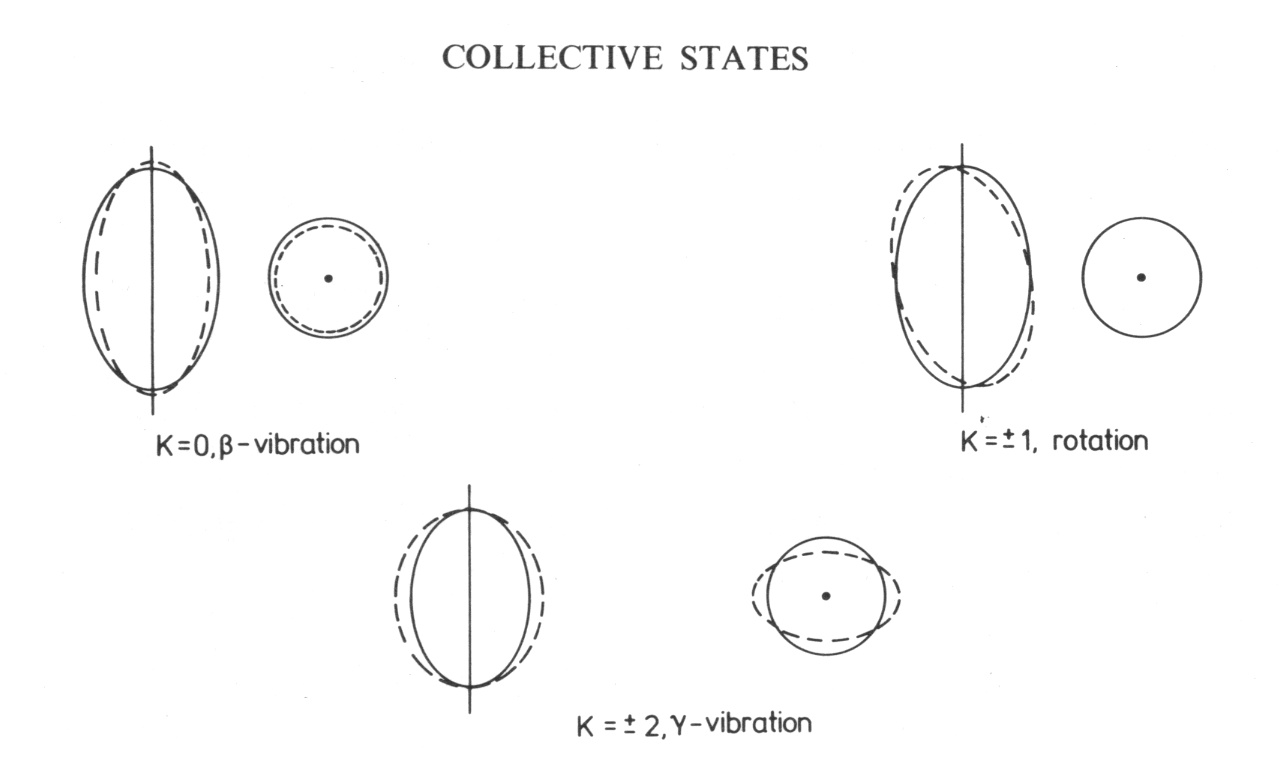}
\end{center}
\caption{\label{fig:1}
Schematic representation of the classical quadrupole shape oscillations in a spheroidal nucleus.}
\end{figure}

The intrinsic wave functions in deformed nuclei with axial symetry are characterized by the parity and the K-quantum number (projection of the angular momentum on the rotational axis). The low-lying $K^\pi$= $0^+$ and $K^\pi$= $2^+$ has been investigated in the past in great detail. In the framework of the incompressible liquid drop model with axially symmetric equilibrium, these low-lying collective states have been interpreted as quadrupole shape oscillations. The $K^\pi$= $0^+$ vibrations which preseve axial symmetry are called $\beta$-vibrations and the $K^\pi$= $2^+$ modes which break the axial symmetry are referred to as $\gamma$-vibrations. In addition one obtains also $K^\pi$= $1^+$ states which correspond to the collective rotation about an axis perpendicular to the symmetry axis. These states are usually called the \emph{spurious} ones, as they are not connected with internal excitations. In microscopic calculations one obtains as many $1^+$ solutions as one has ph components. Only the lowest solution which is the most collective one is "spurious". All the other ones are internal excitations and correspond to real physical states as we will see in section 4.3 section. The major part of the non-spurious (isoscalar) strength is concentrated around 11 MeV. A schematic representation of the quadrupole vibration of the phenomenological liquid drop model are shown in Fig. (\ref{fig:1}). 

In section 4.4 the classical picture is compared with the microscopically calculated transition densities.

\subsection{Low-lying electric states in the rare earth region}

  Experimentally, collective $K^\pi$= $0^+$ excitations with energies of about 1 MeV, the so called $\beta$-vibrations, have been known for a long time. The corresponding $K^\pi$= $2^+$ excitations, the $\gamma$-vibrations, are also present in this energy range. The considerable variations of the energies and transition probabilities over the region of deformed nuclei may cast some doubt on the classical interpretation. Actually, the corresponding high-lying collective states, as we will see, correspond much more that interpretation. The low-lying states are dominated by only a few ph-components and depend therefore sensitively on the single-particle level scheme and the transition densities have little similarity with the classical picture. The theoretical energies and transition probabilities, which can be found in Ref. \cite{Zawischa78}, are in general in fair agreement with the data. 
	
	\subsection{High-lying electric states in the rare earth region}
	
\begin{figure}[htbp]
\begin{center}
\includegraphics[width=12cm]{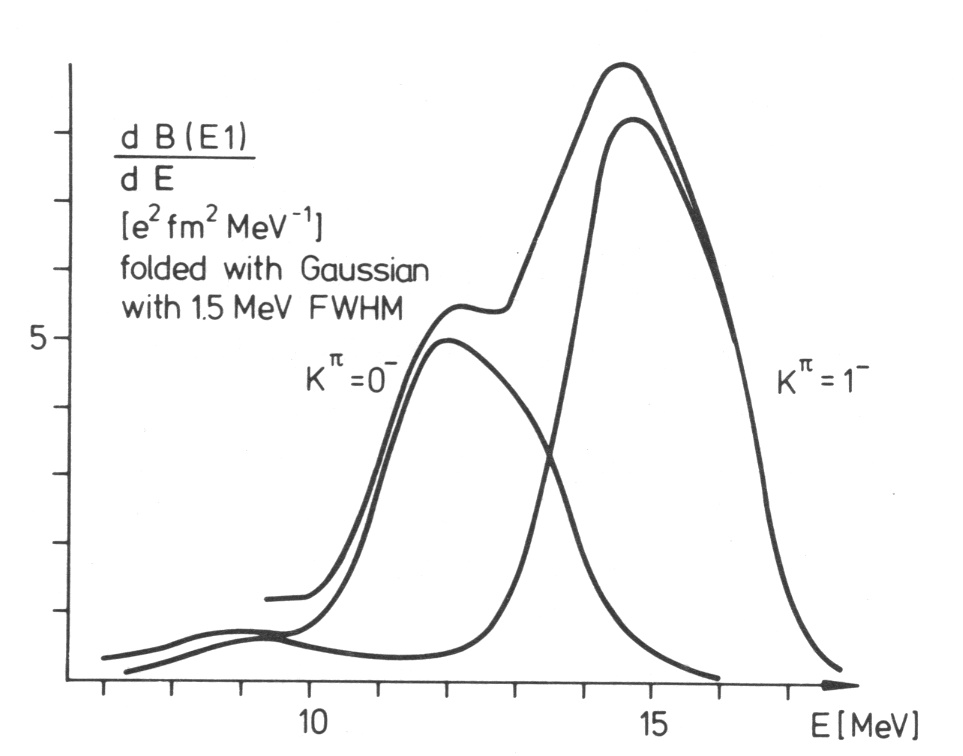}
\end{center}
\caption{\label{fig:3}
Distribution of the $K^\pi$ = $0^-$ and $1^-$ part of the B(E1) transition strength. Here a Gaussian with 1.5 MeV FWHM was folded into the discrete levels.}.
\end{figure}

The high-lying collective states are the well known giant resonances which are qualitatively different in deformed nuclei compared to spherical nuclei. The phenomenological model predicts for a given multipolarity a splitting of the different K-components. This is experimentally well established for the electric dipole resonances \cite{dipole},  where as for the quadrupole resonances a broadening is observed \cite{Harakeh}. From QRPA calculations one obtains in a natural way the excitation energies, transition probabilities and the magnitude of the splitting between different $K^\pi$ components.
\begin{figure}[htbp]
\begin{center}
\includegraphics[width=12cm]{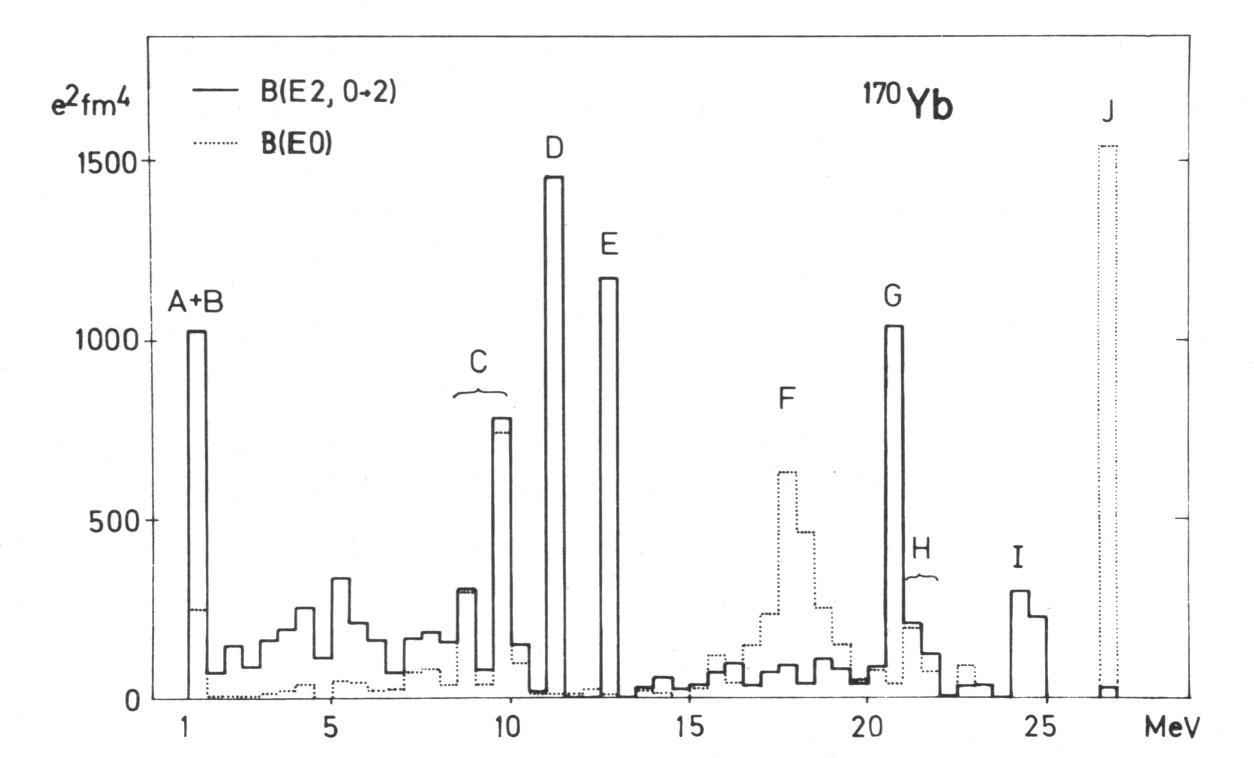}
\end{center}
\caption{\label{fig:2}
Distribution of $B(E2)$ and $B(E0)$ strength in $^{170}$Yb. The $B(E\lambda)$ strength is summed in intervals of 0.5 MeV. The peaks are identified by capital letters. A: lowest $K^\pi$ = $0^+$ excitation. B: lowest $K^\pi$ = $2^+$ excitation. C,D, and E: isoscalar giant quadrupole resonances for $K^\pi$ = $0^+,  1^+$ and $2^+$ components, respectively. F: $K^\pi$ = $0^+$ state ($\Delta T = 0$), predominately of a breathing mode type (the states C and F are both superpositions of $\beta$-vibrations and breathing mode). C,D and F: isovector quadrupole resonances. J: This $K^\pi$ = $0^+$ state ($\Delta$T=1 ) is predominantly an isovector breathing mode.}
\end{figure}
 As an example the results for the $2^+$ and $0^+$ states in $^{170}$Yb are shown in Fig. (\ref{fig:2}). The calculation has been performed in a large, but discrete basis, therefore the theoretical states do not have any width. The $B(E\lambda)$ strength is summed in intervals of 0.5 MeV. An interesting results concerns the isoscalar monopole components (C) and (D) in Fig. (\ref{fig:2}). The $K^\pi$ = $0^+$ of the giant quadrupole resonance (C) possesses also an appreciable monopole strength, therefore also the monopole resonance is split. The calculations for $K^\pi$ = $1^+$ are performed in such a way that the lowest solution is at zero energy. In that case all the spurious strength is concentrated there and the remaining states correspond to internal excitations of the nucleus.
In Fig. (\ref{fig:3}) the result of the electric dipole in $^{170}$Yb is shown. Here a Gaussian with of 1.5 MeV FWHM was folded in all the levels in order to simulate the single particle widths. All results are in fair agreement with the experiments.

\subsection{Microscopic structure of the giant resonances}

\begin{figure}[htbp]
\begin{center}
\includegraphics[width=12cm]{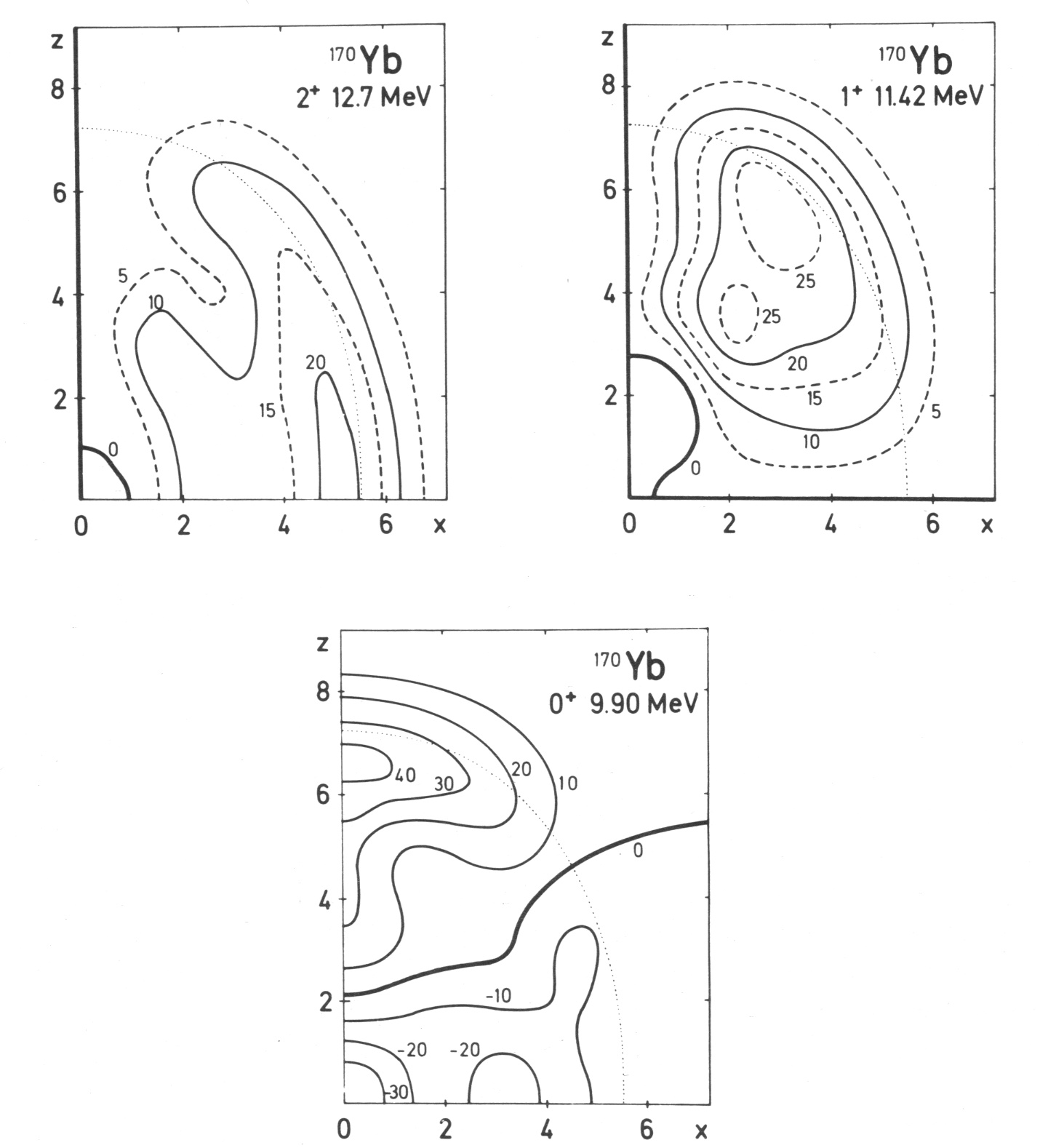}
\end{center}
\caption{\label{fig:4}
Microscopic transition densities between the ground state and the spit K-components of the giant quadrupole resonance in units of $10^{-4}$[$fm^{-3}]$. The dotted line denotes the boundary of the nucleus where the density attains half of his inside value.}.
\end{figure}

In order to get more insight into the nature of the microscopically calculated collective states one may compare the microscopic transition densities with the classical picture. This quantity corresponds most closely to the density change of a classical vibration at maximum elongation. The transition density is defined in the intrinsic coordinate system as: 
\begin{eqnarray}\label{eq:6a}
\rho^{tr}_{K,S}(\mathbf{r} = \sum_{\lambda,\mu}\varphi^*_\lambda(\mathbf{r})\chi^{K,S}_{\lambda,\mu}\varphi_\mu(\mathbf{r})
\end{eqnarray}
where $\varphi_\mu(\mathbf{r})$ and $\chi^{K,S}$ are the single particle wave functions and QRPA amplitudes, respectively; K denotes the corresponding quantum number and S distinguishes the different solutions belonging to the same K. These transition densities have the same symmetries as $Y_{\lambda K}$ and therefore only one quadrant of the z,x plane is drawn in the following examples. As a general result one finds that the transition densities of the giant resonances in different nuclei are very similar all over the deformed rare earth region. Therefore the result for the split giant quadrupole resonance in $^{170}$Yb which is shown in Fig. (\ref{fig:4}) is typical for all well deformed nuclei.
The $K^\pi$ = $2^+$ excitation corresponds closely to a classical $\gamma$-vibration and also for the $K^\pi$ = $1^+$ one gets the expected pattern. The $K^\pi$  = $0^+$ looks very similar to the classical $\beta$-vibration. However, there is also change of the density inside the nucleus, which corresponds to a classical compression mode. As this compression vibration is predominately along the z-axis the authors \cite{Zawischa78} called it \emph{axial breathing mode}. 
The state "F" in Fig. \ref{fig:4} with $K^\pi$ =$0^+$ at 18 MeV corresponds to the breathing mode in a spherical
nucleus. As the corresponding compression vibration \cite{Zawischa78} is perpendicular to the symmetry axis, the authors called it \emph{radial breathing}. This explains why one obtains a splitting of the breathing mode in deformed nuclei.

\section{Isomer Shifts}

An especially nice application of the second order response theory is the calculation of the change of nuclear charge radii due to rotation. These very small effects have been measured by two different methods: (I) applying the M\"ossbauer effect \cite{Wagner78} and (II) using muonic atoms \cite{Goldring71}. In both cases one observes the nuclear $2^+ \rightarrow 0^+$ rotational $\gamma$-transitions in deformed even nuclei. From the view of the liquid-drop model the change of the radii has to be always positive due to the stretching effect. However the experiments showed positive and negative $\delta\left\langle r^2\right\rangle $ in different nuclei, which ruled out this explanation. This paradox, which had been controversially discussed in the seventies, was solved with help of the extended Migdal theory described in Section 2.5.
From the M\"ossbauer experiment the product
\begin{equation}\label{43}
 \delta E^{is}_{Moessbauer}\propto\delta\left|\Psi(0)\right|^2\delta\left\langle r^2\right\rangle  
\end{equation} 
can be extracted, with $\delta\left|\Psi(0)\right|^2$ being the difference of the electron densities at the emitting and the absorbing nucleus and $\delta\left\langle r^2\right\rangle$ the change of the mean-square charge radius

\begin{equation}\label{44}
 \delta\left\langle r^2\right\rangle = \frac{1}{Z}\int d^3\mathbf{r}\; r^{2}_p\;\delta \rho(\mathbf{r})
\end{equation} 
where $\delta \rho(\mathbf{r}) $ denotes the change of the nuclear charge density upon excitation and Z is the charge. In the case of muonic atoms the shift is given as:
\begin{equation}\label{45}
\delta E^{is}_{\mu} = \int d^3\mathbf{r}\; V_\mu(r)\;\delta \rho(\mathbf{r})
\end{equation}
where $V_\mu$ is the Coulomb potential of the muon in the 1s state. As one calculates the change of the density in the second order response one is able to calculate both shifts simultaneously. In muonic atoms one additional complication arises, because the $2^+$ state is split with a strong M1 transition between the two magnetic hyperfine doublets. However, the hyperfine splitting can be calculated within the same theory \cite{Meyer73a}. The authors \cite{Meyer73} calculated the excitation energy of the $2^+$-states, the change of the charge and mass radii and muonic isomerhifts. The results are in a fair agreement with the data. The most remarkable result, however,  is their explanation of the physical mechanism which gives rise to positive and negative $\delta\left\langle r^2\right\rangle $. 
\begin{figure}[htbp]
\begin{center}
\includegraphics[width=8cm]{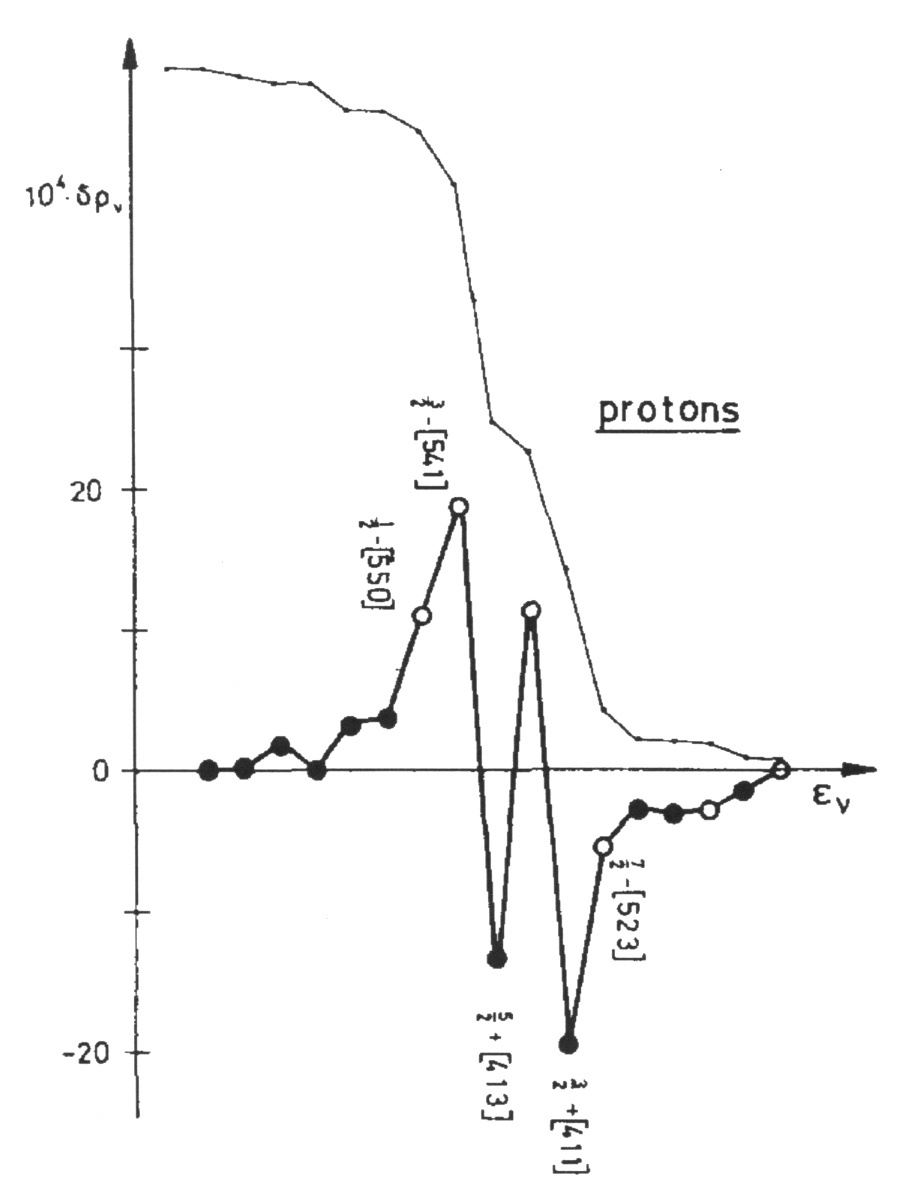}
\end{center}
\caption{\label{fig:5}
The calculated change $\delta \rho_\nu$ of the occupation provabilities of the Nilsson levels near the Fermi energy for protons in $^{152}Sm$. The levels are drawn in a schematic way, equally spaced in the order of increasing energies $\epsilon_\nu$. Open circles refer to N=5, full circles to N=4. The thin line gives the occupation probabilities in the ground state 
(scale is different from $\delta \rho_\nu$}).
\end{figure}
The effect is not connected with the collective motion but with the antipairing effect and the single particle structure near the Fermi edge. The antipairing effect tends to depopulate levels just above the Fermi edge in favor of levels below. It is important to mention that the single particle levels are spit due to the deformation. This has the consequence that near the Fermi edge one has proton states with the main quantum number $N = 4$ and $N = 5$ which have different radii. The change of the radii depends therefore only on a few levels at the Fermi edge.  In Fig. (\ref{fig:6}) the changes of the occupation probabilities of the Nilsson levels for $^{152}$Sm is plotted, where the change of the radius is positive. In $^{160}$Dy the change of the radius is negative and the corresponding changes of the occupation probabilities are shown in Fig.(\ref{fig:7}). 
\begin{figure}[htbp]
\begin{center}
\includegraphics[width=8cm]{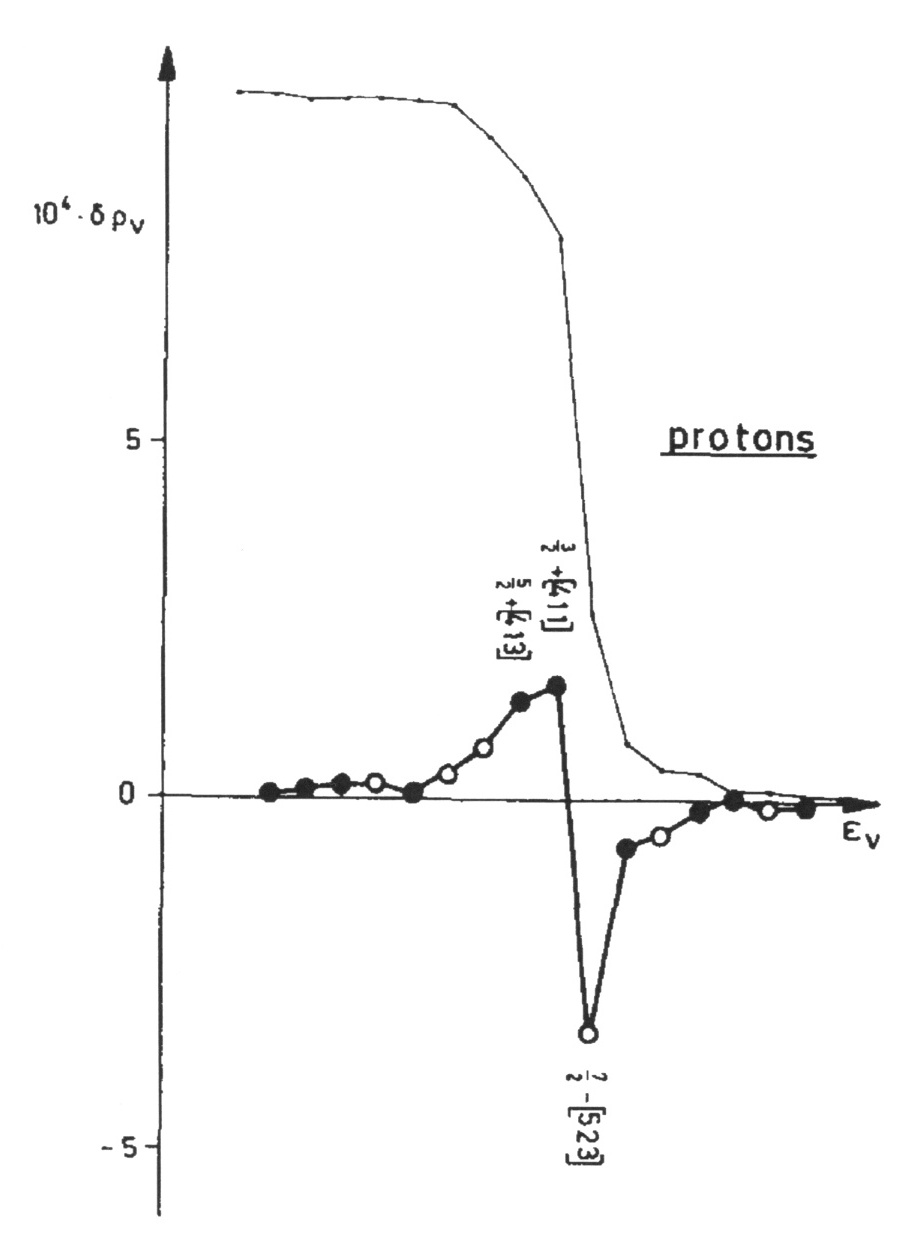}
\end{center}
\caption{\label{fig:6}
Same as in Fig.(\ref{fig:5}), but for $^{160}$Dy.}
\end{figure}
In the case of $^{152}$Sm shown in Fig.(\ref{fig:7}), for three N=5 levels (with larger radii) around the Fermi edge the occupation probabilities are increased and for two N=4 levels they are decreased which result in a positive $\delta\left\langle r^2\right\rangle $. Depopulation of a larger N=5 level in favor of a smaller N=4 level leads to a negative isomer shift. This obviously occurs for $^{160}$Dy due to the ${\frac{7}{2}}{^-}[523]$ proton level just above the Fermi energy. After this very simple explanation, the experimentalists lost the interest in these investigations.

\section{Magnetic Excitations in Deformed Nuclei}
 
Magnetic excitations are calculated in the same way as the electric states discussed in Section 4. There is a large body of experimental data on M1-transitions, that have been reviewed e.g. in \cite{Richter10} and an equally large number of theoretical investigations reviewed by Zawischa \cite{Zawischa98}. There are two different modes: (I) one which is dominated by the orbital transitions and (II) spin-flip transitions which are known from spherical nuclei. While for the first class of transitions the experimental data and the theoretical results are well established, the interpretation of these states in terms of a collective model (scissor modes) has caused some discussions \cite{Zawischa98}. Like in the case of the electric $\beta$- and $\gamma$-vibration of section 4, the low-lying states seem to have less resemblance with the collective model, compared with a predicted high lying, very collective resonance at around 23 MeV \cite{Zawischa89}. 
\begin{figure}[htbp]
\begin{center}
\includegraphics[width=8cm]{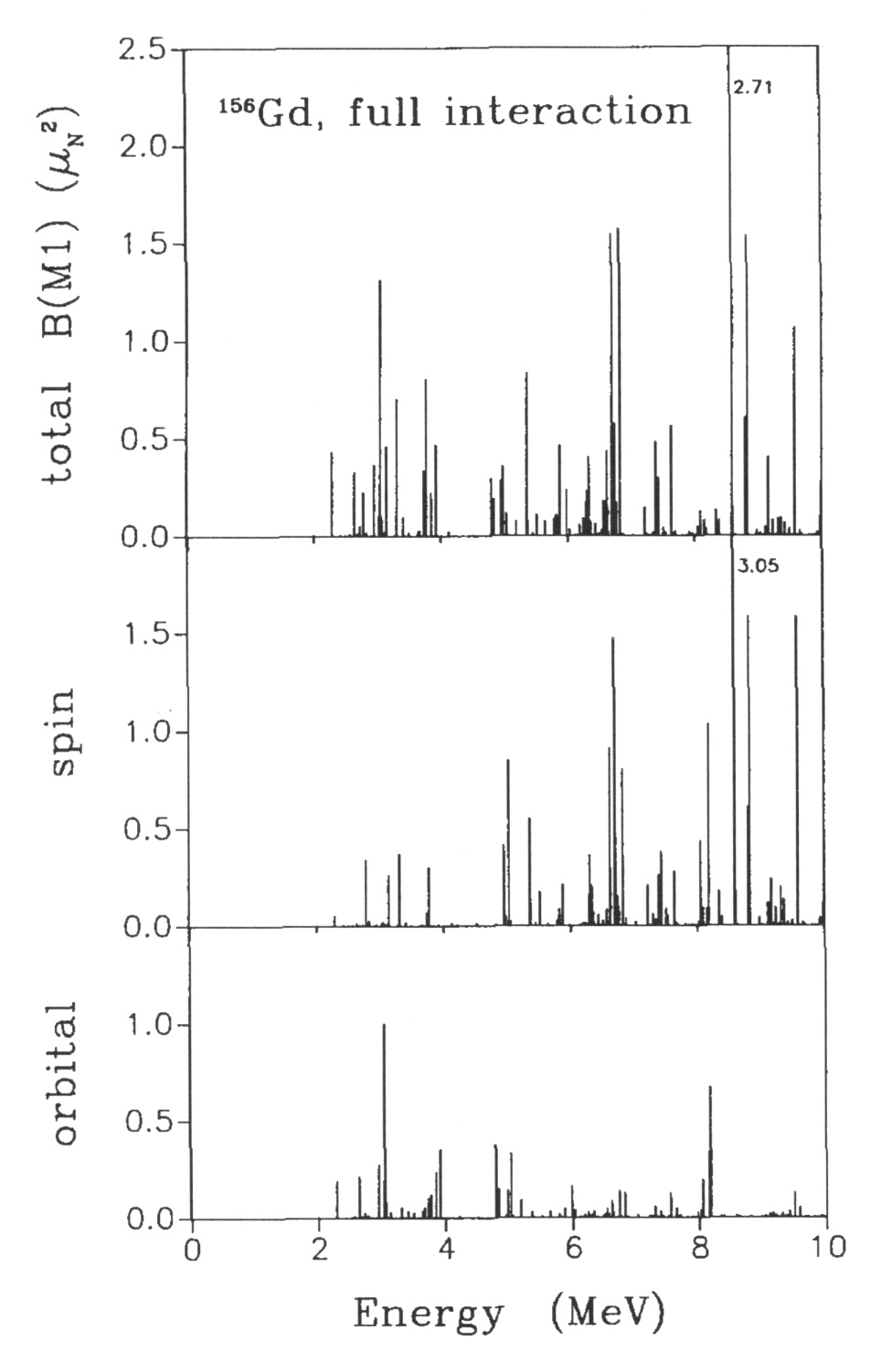}
\end{center}
\caption{\label{fig:7}
Results of a QRPA calculation of the B(M1)$\uparrow$ strength distribution in $^{156}$Gd with the full effective interaction and the bare magnetic operator. Spin-flip and orbital transitions are shown separately.}
\end{figure}
For the spin-flip states such a problem did never appear. The predominant orbital states are energetically lower compared with the spin-flip states. As a typical example, in Fig. (\ref{fig:7}) the results of a QRPA calculation for $^{156}$Gd are presented. Due to the mixture of spin-flip and orbital angular momentum states the Migdal parameter $g^{\prime}_0$ and $f^{\prime}_0$ enter into the calculations. Both ph-force-components are repulsive so that the strength in both cases is shifted to higher energies. 

\begin{figure}[htbp]
\begin{center}
\includegraphics[width=10cm]{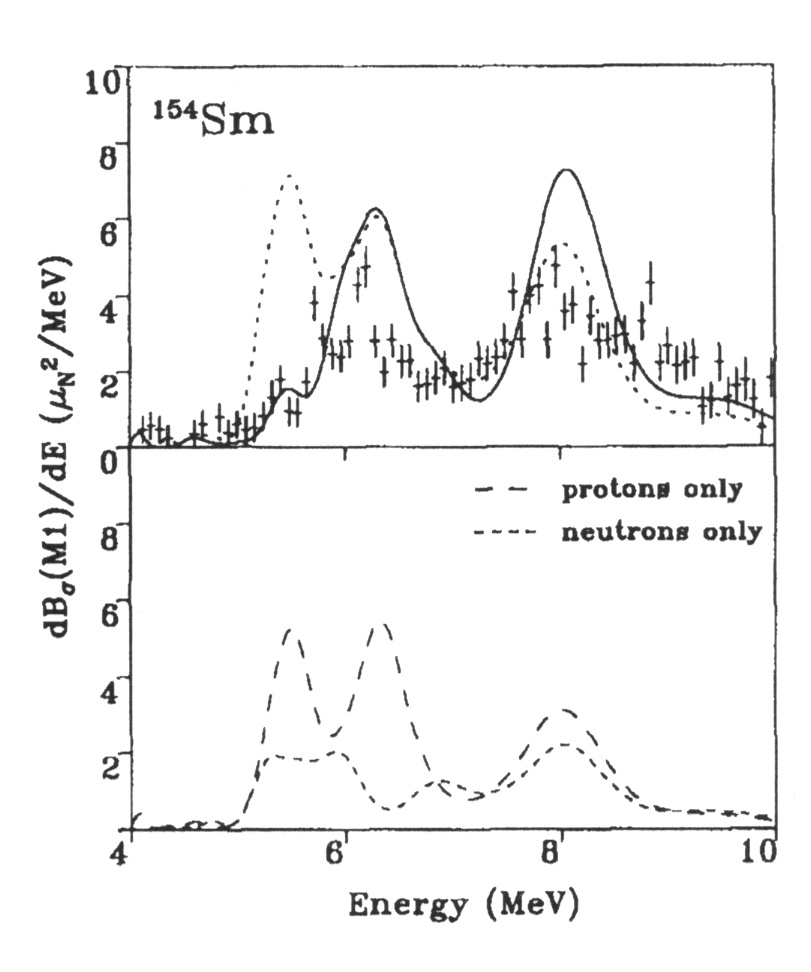}
\end{center}
\caption{\label{fig:8}
Spin-flip strength distribution in $^{154}$Sm. The experimental data\cite{Frekers} are compared with QRPA results.Gaussian have been folded in the theoretical results to produce a continuous distribution.The total theoretical result is given by the full curve. The broken curve in the upper part represents the incoherent sum of the proton and neutron contributions shown in the lower part.}
\end{figure}
There is a clear separation between the orbital components, the so called scissors- modes and the energetically higher spin-flip transitions. 
In Fig. (\ref{fig:8}) a comparison of theoretical results with the data for $^{154}$Sm are given \cite{Zawischa94}. This double bump structure is characteristic for all well deformed rare earth and actinide nuclei \cite{Macfarlane90}. 

\section{Derivation of the Landau parameters}

The Landau-Migdal parameters 
can be derived from an underlying many-body theory of 
nuclear matter and finite nuclei. One has to start from a theory of the nuclear ground state
and perform a functional differentiation of the ground state energy of nuclear matter
 with respect to the quasi particle occupation numbers, as was suggested first by 
G. E.  Brown and S. O. B\"ackmann \cite{Gerry71,Bac68,BJS}.
Several deep insights into nuclear physics have been obtained this way. First of all,
it turned out that the spin-isospin dependent part of the interaction, the famous parameter $g_0'$
could be understood quantitatively from the one-pion exchange and from a reasonable assumption about
short-range correlations. This is ultimately due to the long-range character of the pion-exchange
which allows to treat the effects of short-range correlations in a simplified way by 
essentially removing the short range part of the pion-exchange. 

An important phenomenological generalization of the Landau-Migdal interaction has resulted from 
this observation \cite{Brown80,Stan86}. 
The \emph{Stony Brook Juelich ansatz} augments the conventional Landau-Migdal parametrization 
by the explicit  one-pion and one-rho exchange $V_\pi$ and $V_\rho$ which are folded by a 
correlation function $\Omega_c(\textbf{q})$: 
\begin{equation}\label{eq:6as}
G^{\prime}(q)=\int{\frac{d^3k}{(2\pi)^3}}\left[V_{\pi}(\mathbf{k})+V_{\rho}(\mathbf{k})\right]
\Omega_c(\mathbf{q-k})+\delta{G_{0}^{\prime}}
\mathbf\sigma\cdot\mathbf\sigma^{\prime}\mathbf\tau\cdot\mathbf\tau^{\prime}
\end{equation}
with
\begin{equation}\label{eq:7a}
\Omega_c(\mathbf{q})=(2\pi)^3\delta (\mathbf{q})-\frac{2\pi^2}{q^2}\delta (\left|\mathbf{q}\right|-q_c).
\end{equation}
Here $q_c=3.93$ denotes the inverse of the Compton 
wavelength of the $\omega$ meson. 
A parameter $\delta{G_{0}^{\prime}}$ is introduced to account for a small correction to the
Landau-Migal parameter $G^{\prime}(q)$ that is not produced by the explicit meson-dynamics.

\begin{figure}[htbp]
\begin{center}
\includegraphics[width=10cm]{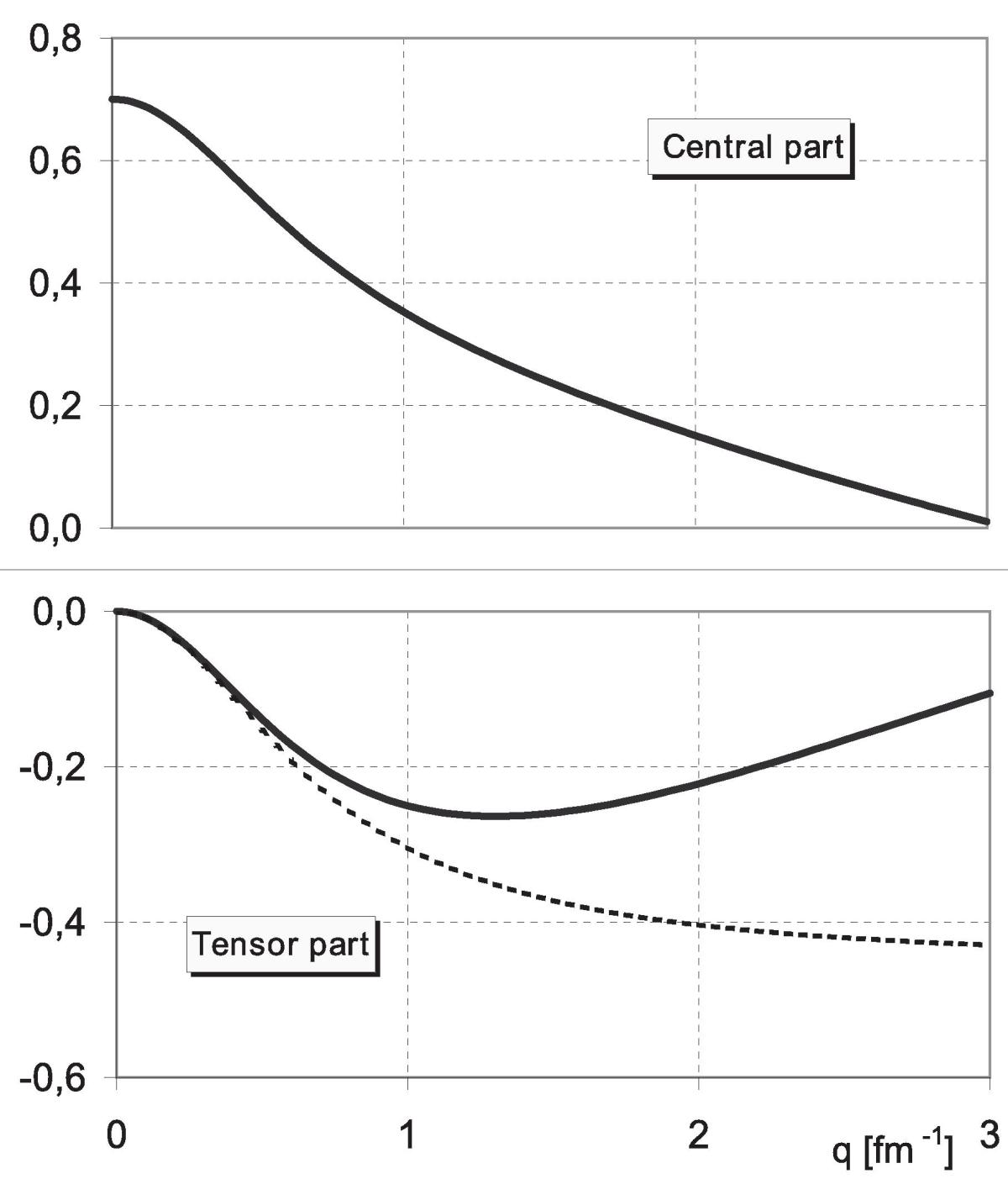}
\end{center}
\caption{\label{fig:2ms}Momentum dependence of the spin-isospin parameter ${G(q)}'$ in units of 
$\left[300 MeV fm^3\right]$. 
The full line is the complete model of Eq. (\ref{eq:6as}), the dashed line in the lower part is the 
correlated one-pion exchange only.}
\end{figure}

In the upper part of Fig. (\ref{fig:2ms}) the q-dependence of $G'$ is plotted. 
One realizes that at small momentum transfers the central part of the spin-isospin interaction is strongly 
repulsive and for larger momentum transfers it is weak. 
The central part  of the $\pi$ -meson and $\rho$ -meson contributions have the same sign, whereas  the 
tensor parts have the opposite signs. 
The $\rho$ -meson exchange therefore acts as a natural cut-off for the strong tensor component of 
the one-pion exchange. 
All details of the calculations are given in the original publications \cite{Brown80}.

\begin{figure}[htbp]
\begin{center}
\includegraphics[width=10cm]{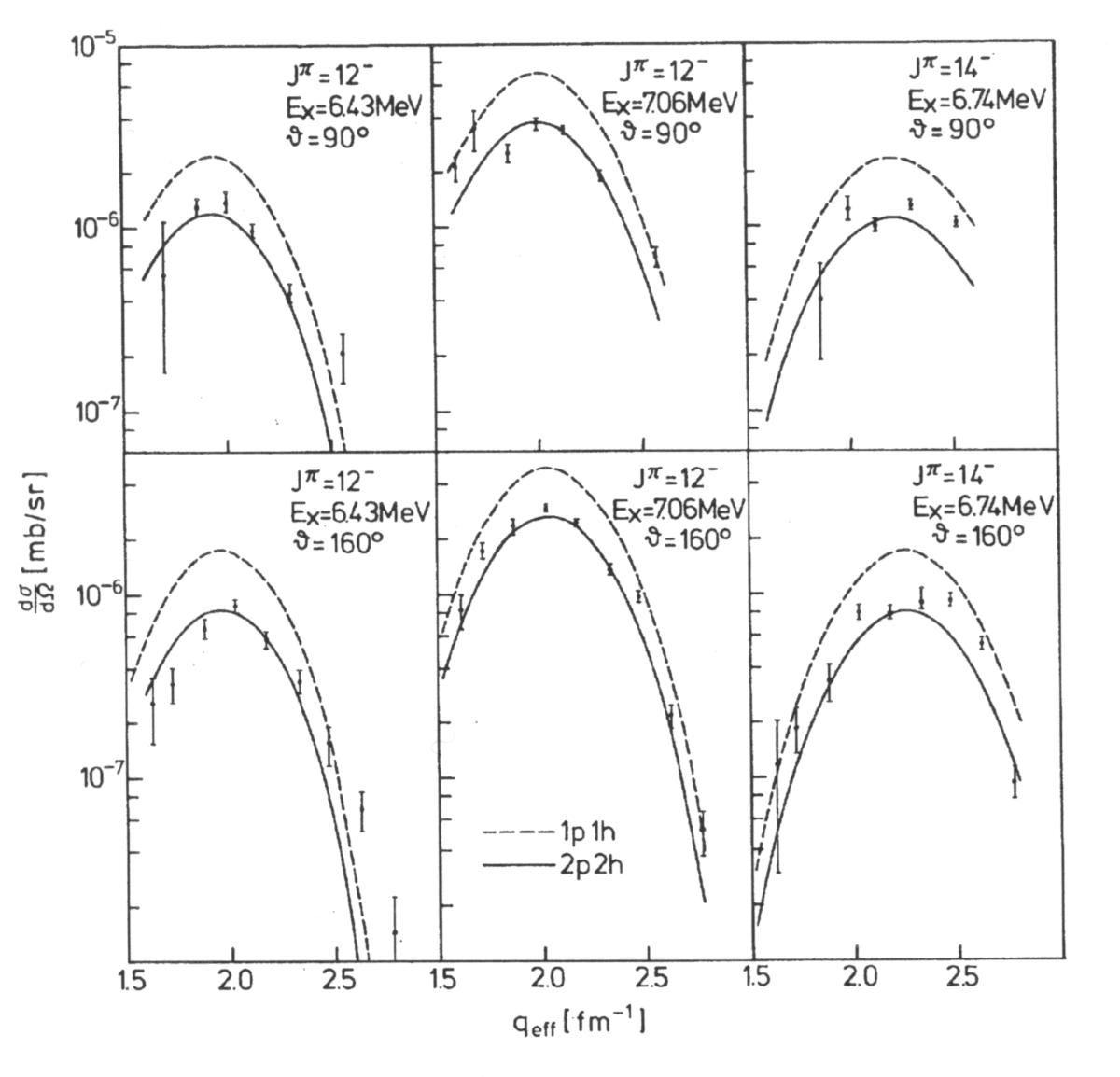}
\end{center}
\caption{\label{fig:8s} Inelastic electron scattering cross sections at $\Theta = 90\deg$ and $\theta = 160\deg$ 
of the magnetic high spin states in $^{208}$Pb. Experiments are compared with the RPA results (dashed lines) and the extended model. Calculations have been done in DWBA.}
\end{figure}

The unnatural parity $12^-$ and $14^-$ magnetic high spin states 
discovered experimentally by Jochen Heisenberg and his collaborators at the BATES electron scattering facility \cite{Heiss78} 
in $^{208}Pb$ are a striking example to illustrate the momentum dependence of the
generalized Landau-Migdal interaction.
The cross sections peak around $q\approx 2\; [fm^{-1}]$, see Fig. (\ref{fig:8s}).
The energies of the two $12^-$ and the one $14^-$ states are close to the experimental ph-energies but the 
cross sections are only half of the shell model prediction. 
The explanation for this surprising result was given in Ref. \cite{Krewald80}. 
As the spin-isospin interaction in the relevant momentum range is essentially zero, the RPA solutions are 
close to the unperturbed ph-energies. 
In their extended model the authors included also the effects of the low-lying phonons within the 
so called \emph{core coupling RPA} which provides an explanation for the reduction of the cross sections.

 While magnetic modes with high angular momenta do not show collectivity,  magnetic modes of small multipolarities may show
surprisingly large cross sections.
The most outstanding mode of this kind is the Gamov-Teller resonance discovered by Charles Goodman and Dan Horen
 at the Indiana cyclotron facility in charge-exchange reactions \cite{Charles80,PL}. 
 In Tab. (\ref{Tab:1}), the averaged ph-energies $\epsilon_{ph}$ are compared with the 
RPA excitation energies E$_{RPA}$ for  
the $0^-$, $1^-$ $1^+$ and $2^-$ spin-isospin modes in $^{208}$Pb.
In the first three cases, the energy shift $\Delta$E is of the order of 5 MeV. 
The $2^-$ result is qualitatively different, however.
Here the ph-force is weak as the transition density is peaked at larger momentum transfer, and as a consequence one obtains four states 
which are only little shifted from the uncorrelated ph-energies. 
The comparison with the data shows a fair agreement as far as the mean energies are concerned. 

\begin{table}[ht]
\begin{tabular}{rrrrr}
     J$^\pi$ & $\epsilon_{ph}$ [MeV] & $E_{RPA}$ [MeV] &    $\Delta E$ [MeV] & $E_{exp}$ [MeV] \\
\hline
       $0^-$ &       21.8 &       26.6 &        4.8 & 25.1 $\pm$ 1.0 \\
\hline
       $1^+$ &       13.1 &       18.9 &        5.8 & 19.2 $\pm$ 0.2 \\
\hline
       $1^-$ &       21.5 &       26.3 &        4.8 & 25.1 $\pm$ 1.0 \\
\hline
             &       20.7 &       22.6 &        1.9 &            \\

       $2^-$ &       20.1 &       21.2 &        1.1 & 25.1 $\pm$ 1.0 \\

             &       23.3 &       24.5 &        1.2 &            \\

             &       28.0 &       28.8 &        0.8 &            \\
\hline
\end{tabular}
\caption{Charge-exchange resonances in $^{208}$Pb.}
\label{Tab:1}
\end{table}

\begin{figure}[htbp]
\begin{center}
\includegraphics[width=10cm]{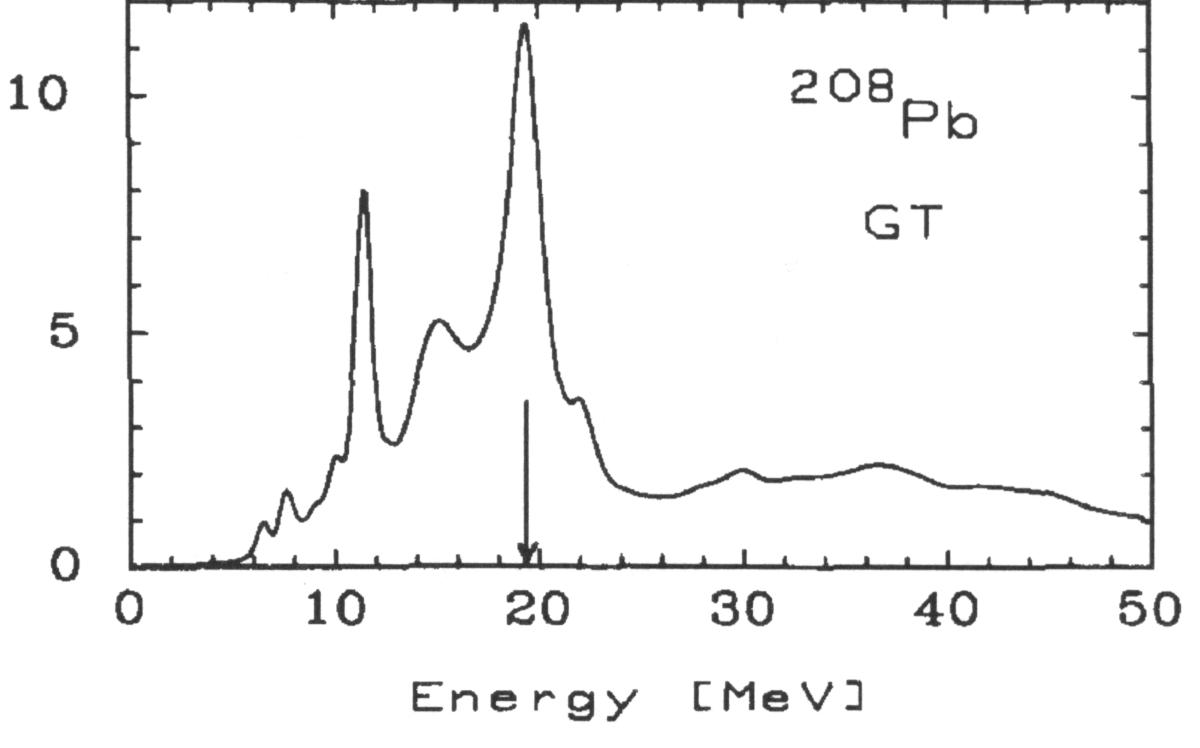}
\end{center}
\caption{\label{fig:6s} Gamow-Teller resonance in $^{208}$Pb calculated in an extended RPA model \cite{Stan86}. The arrow indicated the experimental mean energy.}
\end{figure}

The charge exchange resonances have relatively large widths which can not be obtained within a 1p1h approach 
but one has to include higher configurations. 
In Fig. (\ref{fig:6s}) an example for such more involved calculations is shown. 
The authors \cite{Stan86} extended the conventional RPA approach and include 1p1h as well as 2p2h-configurations 
in a consistent way. 
This calculation not only reproduced the known experimental energy and width but it also predicted a 
long tail up the 50 MeV, where nearly half of the strength is hidden. 
A detailed discussion of spin-isosopin modes is given in Ref. \cite{O92}.

In the early 1970's, there were speculations about the existence of a pion condensate in nuclei. At large
nuclear densities, the spins of the pions were supposed to align as a consequence of the one-pion exchange
interaction. The investigation of high energy heavy ion collisions were suggested as an experimental tool
to study the supposed phase transitions, despite warnings that the time-scale for performing a phase
transition does not match with the time two high energy ions need to pass each other \cite{krew80}.
More over, one had to
realize that the pion-exchange is always accompanied by short-range
 nuclear correlations which generate a Landau-Migdal parameter $g_0'$,
and thus  suppress the onset of a phase transition \cite{krew80}.

In the case of the Landau parameter $f_0$, microscopic derivations had to face severe challenges.
Landau had related the
parameter $f_0$ to the compression modulus of nuclear matter,
$$ K = 3 \frac { \hbar^2 k_F^2} {m} (1 + f_0).$$
Inserting the empirical  value $K = 210 \pm 30$MeV, one finds that $f_0$ cannot be negative.
Moreover, the isotope shifts in heavy nuclei clearly rule out negative values of $f_0$ \cite{Speth77}.

Brueckner's theory of nuclear matter
produces values of the
spin- and isospin- independent parameter violating Landau's stability criterion 
 $f_0 > -1,$ see Table (\ref{tab:landau-param}).

\begin{table}
  \caption{
Landau parameters of nuclear matter for a Fermi momentum of $k_F = 1.36 $fm$^{-1}$,
obtained by using the one-boson exchange potential HEA \cite{HEA}
 as the
 two-nucleon interaction and different versions of Brueckner theory to derive the
binding energy of nuclear matter. First row: standard Brueckner Hartee-Fock.
Second row: The Dirac spinors of the bare two-nucleon interaction are modified
 Third row:
the empirical parameters determined from linear response of finite nuclei.
All parameters are given in units of 300 MeV fm$^3$.
}
\begin{center}
\begin{tabular}{lllll}
 \hline
method			&$f_0$	&$f_0'$ 	&$g_0$ & $g_0'$	  \\
 \hline
BHF, non-relativistic	& -1.17		& 0.32		& 0.2 &		0.63       \\
BHF, relativistic	& -0.72		& 0.42		& 0.12 &	0.63       \\
empirical               & $+0\pm 0.2$	& 0.8		& 0.2   &	0.9       \\
\hline\hline
\end{tabular}
\end{center}
\label{tab:landau-param}
\end{table}  
	
This is an expression of the fact that Brueckner theory does not saturate nuclear
matter at the empirical saturation   point, but at much larger densities. The advent of Walecka's
relativistic mean field theory suggested a new mechanism to saturate nuclear matter at the correct saturation point \cite{Serot:1984ey}.
Indeed, incorporating the relativistic effects suggested by Walecka into a Brueckner calculation,
one is able to produce a Landau-Migdal parameter $f_0$ which signals stability of the nuclear ground state,
but quantitatively, it does not agree with the empirical one.
This clearly shows that the Walecka approach is incomplete \cite{Krew88}. 
A hint to the relevance of three-body interactions came from phenomenology.
 The effective interactions of the
Skyrme type rely on medium-dependent effective interactions which may be interpreted as effects due to three-body
forces. The different phenomenological Skyrme-like parameterizations allow a dramatic variation in the
resulting Landau-Migdal parameters, however, which blocked progress. Indeed, instead of deriving the
Landau parameters from an underlying many body approach, one rather used the empirical 
 Landau-Migdal parameters to constrain the  freedom available in the choice of appropriate
parameterizations of the effective medium dependent interactions \cite{liu91}.
For progress, a theory of the three-nucleon interactions was required.
In the 1980's the then new field of hadron physics emerged and improved our understanding of
hadronic interactions. Effective Field Theories for pion-pion scattering, pion-nucleon interactions,
and eventually nucleon-nucleon scattering were developed and provided a systematic expansion scheme
which restricts the vast number of three-nucleon interactions allowed in older theories. 
A recent summary is given in Ref. \cite{EHM08}.
 In 
next-to-next-to-leading order, there are exactly three types of three-nucleon interactions.
First theoretical investigations of nuclear matter based on those interactions have become available
and are expected to make a major impact on the theoretical command of the Landau-Migdal
 parameters \cite{Bogner,Holt}.
A successful description of neutron matter based on chiral three-nucleon interactions has been published and promises
a solid basis for the theory of neutron-rich nuclei \cite{Hebeler2010}.

\section{Conclusion}
We have reviewed the application of the conventional \emph{TFFT} to deformed nuclei in the rare earth and actinide region. The electric as well as magnetic states have been calculated. The quasi particle in the definition of Landau are the Nilsson single particle levels. Whereas in spherical nuclei the corresponding single particle levels are  $2j+1$ times degenerate, this degeneracy is lifted due to the deformation. For that reason the density of single particle states is much higher and one does not need e.g. an additional splitting due to low-lying collective states. The theoretical results are in general in good agreement with the experimental data. The low-lying $\beta$- and  $\gamma$- vibrations are not very collective, therefore the theoretical results depend strongly on the single particle energies. We also reviewed an extension of Migdal's theory, the so called second order response theory. An especially nice example are the isomer shifts of rotational states which we discussed in some details.
Finally we discussed more recent developments in the frame work of effective theories.

\section{Acknowledgement}
It is a pleasure to thank S. Dro\.{z}d\.{z}, V. Klemt, J. Meyer-ter-Vehn, P. Ring, G. Tertychny$^+$, J. Wambach and D. Zawischa for the collaboration in developing and extending the Landau-Migdal theory in the past, S. Avdeenkov, S. Kamerdziev, N. Lyutorovich and V. Tselyaev for their collaboration up to present. We thank the Deutsche Forschungsgemeinschaft for the grants DFG: 436 RUS 113/994/0-1 and DFG: 436 RUS 113/806/0-1. JS thanks the Foundation for Polish Science for financial support through the Alexander von Humboldt Honorary Research Fellowship.
\appendix


\begin{thebibliography}{99}
\bibitem{Migdal67} A.B. Migdal, \textit{Theory of Finite Fermi Systems and Application to Atomic Nuclei}, 
                   Wiley, New York, 1967.
\bibitem{Speth77} J. Speth, E. Werner and W. Wild, Physics Reports \textbf{33}, (1977) 127. 
\bibitem{Speth70} J. Speth Z. Physik \textbf{239} (1970) 249.             
\bibitem{Ring74} P. Ring and J. Speth, Nucl. Phys. \textbf{A235}, (1974) 315.                       
\bibitem{KS82} V. A. Khodel and E. E. Saperstein Physics Reports \textbf{92}, (1982) 183.
\bibitem{Stan90} S. Dro$\dot{z}$d$\dot{z}$, S. Nishizaki, J. Speth and J. Wambach, 
                 Physics Reports \textbf{197}, (1990) 1.
\bibitem{rev04} S. Kamerdzhiev, J. Speth and G. Tertychny, Phys. Rep. \textbf{393}, (2004) 1.
\bibitem{Tselyaev89} V.I. Tselyaev, Sov.J.Nucl.Phys. \textbf{50}, (1989) 780.                 
\bibitem{Tselyaev07} V.I. Tselyaev, Phys. Rev. \textbf{C75}, (2007) 024306. 
\bibitem{GS06} F. Gr\"ummer and J. Speth J. Phys. G: Nucl. Part. Phys. \textbf{32} (2006) R192.
\bibitem{Krewald09} S. Krewald and J. Speth, Int. Journal of Mod. Phys.\textbf{18} (2009) 1425.
\bibitem{TsSp07} V.I. Tselyaev, J. Speth, F. Gr\"ummer, S. Krewald, A. Avdeenkov, E.V. Litvinova and G. Tertychny,  
                Phys. Lett. \textbf{B653}, (2007) 196. 
\bibitem{Ring04} D.Vretenar, N. Paar, P. Ring and G.A. Lalazissis, Nucl.Phys. \textbf{A692}, (2004) 281. 
\bibitem{LiTselyaev07} E.V. Litvinova and V.I. Tselyaev, Phys. Rev. \textbf{C75}, (2007) 054318.
\bibitem{Ring01} D.Vretenar, T. Niksic, N. Paar and P. Ring,  Nucl.Phys. \textbf{A731}, (2001) 496.
\bibitem{Lenske02} N.Ryezayeva et al. Phys. Rev. Lett. \textbf{89}, (2002) 2727502.            
\bibitem{Ring02} D. Vretenar et al.Phys. Rev. \textbf{C65}, (2002) 021301.  
\bibitem{Krew88} S. Krewald, K. Nakayama and J. Speth, Phys.Rep. \textbf{161} (1988) 103.
\bibitem{Baym62} L. P. Kadanoff and G. Baym, \textit{Quantum statistical mechanics}, W. A. Benjamin,Inc, New York, 1962.   
\bibitem{Wagner63} W. Brenig and H. Wagner, Z. Physik \textbf{173} (1963) 484. 
\bibitem{Meyer73} J. Meyer and J. Speth, Nucl. Phys. \textbf{A203} (1973) 17.
\bibitem{Urin1} M. G. Urin and D. Z. Zaretsky, Nucl.Phys. \textbf{35} (1961) 219.
\bibitem{Urin2} M. G. Urin and D. Z. Zaretsky, Nucl.Phys. \textbf{47} (1963) 97.
\bibitem{Bir73} B. L. Birbrair, Nucl.Phys. \textbf{A212} 27.
\bibitem{Landau}  L. D. Landau, JETP \textbf{3} (1957) 920; \textbf{5} (1957) 101; \textbf{8} (1959) 70.
\bibitem{Larkin63} A. L. Larkin and A. B. Migdal, JETP(Sov.Phys.)\textbf{17} (1963) 1146.  
\bibitem{Birbrair} B. L. Birbrair, Nucl. Phys. \textbf{A108} (1968) 449.
\bibitem{Sergey69} S. P. Kamerdzhiev, Sov. J. Nucl. Phys. \textbf{9} (1969) 190.
\bibitem{Speth69} J. Speth, Nucl. Phys. \textbf{A135} (1969) 445. 
\bibitem{Gorkov} L. P. Gorkov JETP(Sov.Phys.)\textbf{7} (1958) 505.             
\bibitem{Bogolyubov59} N. N. Bogolyubov, ZhETF(UDSSR) \textbf{67} (1958) 58. 
\bibitem{Baranger60} M. Baranger, Phys.Rev.\textbf{120} (1960) 957.    
\bibitem{Belyaev} S. T. Belyaev, Nucl. Phys. \textbf{64} (1965) 17.  
\bibitem{MS72} J.Meyer and J. Speth, Phys. Lett. \textbf{B39} (1972) 330. 
\bibitem{Migdal59} A. B. Migdal, Nucl. Phys. \textbf{13} (1959) 655.
\bibitem{Bir68} B. L. Birbrair, Nucl. Phys. \textbf{A108} (1968) 449.
\bibitem{Meyer72} J. Meyer, J. Speth and J. H. Vogeler, Nucl. Phys. \textbf{A193} 1972 60.
\bibitem{BM} A. Bohr and B. Mottelson, Nuclear Structure, Vol.II (Benjamin, reading, Mass., 1975)
\bibitem{Vogeler} J. H. Vogeler, Nucl.Phys. \textbf{A133} (1969) 289.
\bibitem{Gustaf} C. Gustafson, I. L. Lamm, B. Nilsson and S. G. Nilsson, Ark. Fys. \textbf{36} (1971) 69.
\bibitem{Zawischa78} D. Zawischa, J. Speth and D. Pal, Nucl.Phys. \textbf{A311} (1978) 445.
\bibitem{dipole} E. G. Fuller and E. Hayward, Nucl.Phys. \textbf{30} (1962) 613.
\bibitem{Harakeh} M. N. Harakeh and A. van der Woude, \textbf{Giant Resonances}, Oxford University Press, Oxford, 2001.
\bibitem{Wagner78} G. K. Shenoy and F. W. Wagner, \textit{ M\"ossbauer Isomer Shifts} North-Holland, Amsterdam, 1978.
\bibitem{Goldring71} G. Goldring and R. Kalish, \textit{Hyperfine Interaction in Excited Nuclei}, Gordon and Breach, New York, 1971.
\bibitem{Meyer73a} J. Meyer, P. Ring and J. Speth, Phys. Rev. \textbf{C7} (1973) 1803.
\bibitem{Richter10} K. Heyse, P.v. Neumann-cosel and A. Richter, Rev. Mod. Phys. \textbf{82} (2010) 2365
\bibitem{Zawischa98} D. Zawischa, J.Phys. G:Nucl.Part.Phys. \textbf{24} (1998) 683.
\bibitem{Zawischa89} D. Zawischa and J. Speth, Phys.Lett. \textbf{219B} (1989) 529.
\bibitem{Zawischa94} D. Zawischa and J. Speth, Nucl.Phys. \textbf{A569} (1994) 343c.
\bibitem{Macfarlane90} D. Zawischa, M. Macfarlane and J. Speth, Phys. Rev. \textbf{C42} (1990) 1461.
\bibitem{Frekers} D. Frekers et al., Phys.Lett. \textbf{B244} (1990) 178.
\bibitem{Gerry71}  G.E. Brown, Rev. Mod. Phys. \textbf{43} (1971) 1.
\bibitem{Bac68} S.O. B\"ackmann, Nucl. Phys. \textbf{A120} (1968) 593.
\bibitem{BJS} S.O. B\"ackman, A. D. Jackson and J. Speth, Phys.Lett. \textbf{B56} (1975) 209.
\bibitem{Brown80} J. Speth, W. Klemt, J. Wambach and G. E. Brown, Nucl.Phys. \textbf{A343} (1980) 382.
\bibitem{Stan86} S. Dro\.zd\.z, V. Klemt, J. Speth and J. Wambach Phys.Lett.  \textbf{166B} (1986) 18.
\bibitem{Heiss78} J. Lichtenstadt, J. Heisenberg, C.N.Papanicolas, C.P. Sarget, A.N. Courtemanche and J.S. McCarty, Phys.Rev.Lett.\textbf{40} (1978) 1127.   
\bibitem{Krewald80} S. Krewald and J. Speth, Phys.Rev.Lett. \textbf{45} (1980) 458.
\bibitem{Charles80} C. D. Goodman et al. Phys.Rev.Lett.\textbf{44} (1980) 1755.   
\bibitem{PL}  W.G.Love, in \textit{The (p,n)Reaction and the Nucleon-Nucleon Force}, edited by C.D. Goodman et al. Plenum, New York, 1980, p. 30; F. Petrovich, \textit{ibid.}, p. 135.  
\bibitem{HEA} K. Holinde, K. Erkelenz, and R. Alzetta, Nucl. Phys. \textbf{A198} (1972) 598.
\bibitem{O92} F. Osterfeld, Rev. Mod. Phys.\textbf{64} (1992) 491.
\bibitem{krew80} S. Krewald and J. W. Negele, Phys.Rev.\textbf{C21} (1980) 2385.
\bibitem{Serot:1984ey}  B.D. Serot and J.D. Walecka,  Adv.\ Nucl.\ Phys.\  {\bf 16} (1986) 1.
\bibitem{liu91} K. F. Liu, H. Luo, Z. Ma, Q. Shen, Nucl. Phys. \textbf{A534} (1991) 1.
\bibitem{EHM08} E. Epelbaum, H.-W. Hammer, and Ulf-G. Mei{\ss}ner, Rev. Mod. Phys. \textbf{81}(2009) 1773.
\bibitem{Bogner}S. K. Bogner, R. J. Furnstahl, A. Schwenk, Prog. Part. Nucl. Phys. \textbf{65} (2010) 94.
\bibitem{Holt}J. W. Holt, G.E. Brown, J. D. Holt, and T. T. S. Kuo, Nucl. Phys. \textbf{A785} (2007) 322.
\bibitem{Hebeler2010} K. Hebeler and A. Schwenk, Phys. Rev. \textbf{C82} (2010) 014314.

\end{thebibliography}
\end{document}